\documentclass[runningheads]{llncs}

\usepackage{graphicx} 
\RequirePackage{xcolor}
\usepackage[hidelinks]{hyperref}
\usepackage{siunitx}
\usepackage{xurl}
\usepackage{booktabs}
\usepackage{tikz-network}
\usepackage[most]{tcolorbox}
\usepackage{makecell}
\usepackage{bm}
\usepackage{multirow}
\usepackage{standalone}
\usepackage{amssymb}
\usepackage{array}
\usepackage{ragged2e}
\usepackage{longtable}
\usepackage{booktabs}

\usetikzlibrary{shapes.misc}
\tikzset{cross/.style={cross out, draw=black, minimum size=2*(#1-\pgflinewidth), inner sep=0pt, outer sep=0pt},
	cross/.default={1pt}}

\def\ShiftsProposalN/{\num{1202}}
\def\ShiftsProposalPct/{\num{14.81}}
\def\ShiftsProposalMax/{\num{30}}
\def\ShiftsProposalMedian/{\num{1.0}}
\def\ShiftsProposalStd/{\num{2.64}}
\def\ShiftsSpaceN/{\num{229}}
\def\ShiftsSpacePct/{\num{64.15}}
\def\ShiftsSpaceNHighTvlM/{\num{30}}
\def\ShiftsSpaceNHighTvlB/{\num{2}}
\def\ShiftsDAOContributorN/{\num{1457}}
\def\ShiftsContributorN/{\num{1362}}
\def\ShiftsContributorPct/{\num{3.37}}
\def\ShiftsContributorsProposalN/{\num{728}}
\def\ShiftsContributorsProposalPct/{\num{60.57}}
\def\ContributorsN/{\num{17561}}
\def\SpaceHasContributors/{\num{12272}}
\def\SpaceContributorsAvg/{\num{1.43}}
\def\ContributorComponents/{\num{10651}}
\def\ProjGDCSpaceAvgDegree/{\num{0.64}}
\def\SpaceAssociateVotes/{\num{794}}
\def\SpaceSelfAssociates/{\num{529}}
\def\ProposalChoiceScoreNotMatch/{\num{17}}
\def\ProposalNotFinal/{\num{1806}}
\def\ProposalNotFinalPerc/{\num{3.84}}
\def\ProposalSingleCoicePct/{\num{86.97}}
\def\SpacesSelfAssociatesDecided/{\num{178}}
\def\SpacesSelfAssociatesDecidedPct/{\num{20.41}}
\def\ProposalsSelfAssociatesDecided/{\num{2100}}
\def\ProposalsSelfAssociatesDecidedPct/{\num{5.98}}
\def\ProposalsNoneSelfAssociatesDecided/{\num{1847}}
\def\SpacesToValidate/{\num{367}}
\def\VotesToValidate/{\num{473306}}
\def\VotesConsistency/{\num{461402}}
\def\VotesConsistencyPct/{\num{97.48}}
\def\ProposalsTimeOut/{\num{0}}
\def\SpacesRaw/{\num{12294}}
\def\ProposalsRaw/{\num{76851}}
\def\VotersRaw/{\num{1603994}}
\def\VotesRaw/{\num{8365707}}
\def\ContributionRaw/{\num{11949}}
\def\ContrVotesRaw/{\num{316900}}
\def\SpacesClean/{\num{872}}
\def\ProposalsClean/{\num{35124}}
\def\VotersClean/{\num{986557}}
\def\VotesClean/{\num{5240622}}
\def\ContributionClean/{\num{7478}}
\def\ContrVotesClean/{\num{191507}}
\def\SpacesValidated/{\num{357}}
\def\ProposalsValidated/{\num{8116}}
\def\VotersValidated/{\num{119413}}
\def\VotesValidated/{\num{438668}}
\def\ContributionValidated/{\num{3927}}
\def\ContrVotesValidated/{\num{22878}}
\def\SpacesValidatedPct/{\num{40.94}}
\def\ProposalsValidatedPct/{\num{23.11}}
\def\VotersValidatedPct/{\num{12.1}}
\def\VotesValidatedPct/{\num{8.37}}
\def\ContributionValidatedPct/{\num{52.51}}
\def\ContrVotesValidatedPct/{\num{11.95}}
\def\SpacesNVPmoreOne/{\num{580}}
\def\SpacesNVPmoreFifty/{\num{66}}
\def\SpacesNVPmoreFiftyPct/{\num{7.54}}
\def\SpacesNVPmoreTen/{\num{297}}
\def\SpacesNVPmoreTenPct/{\num{33.94}}
\def\SpacesNVPmoreHundred/{\num{9}}
\def\VpMedian/{\num{4.26}}
\def\VpStd/{\num{21.22}}
\def\StrategiesN/{\num{208}}
\def\StrategiesErcPct/{\num{53.85}}
\def\SpacesOrigN/{\num{9505}}
\def\ProposalsOrigN/{\num{76851}}
\def\VotersOrigN/{\num{1603994}}
\def\SpacesSingleProposalPct/{\num{42.67}}
\def\SpacesFollowersLessFivePct/{\num{54.05}}
\def\SpacesLessTwoVotesPct/{\num{50.85}}
\def\SpacesSelectedN/{\num{967}}
\def\ProposalsSelectedN/{\num{47037}}
\def\VotesSelectedN/{\num{7193060}}
\def\ProposalsSelectedPct/{\num{61.21}}
\def\VotesSelectedPct/{\num{87.61}}

\definecolor{contributor_red}{RGB}{162,20,0}

\newtcbtheorem{Summary}{\bfseries Summary}{enhanced,drop shadow={black!90!white},
	coltitle=black,
	top=0.3in,
	attach boxed title to top left=
	{xshift=1.5em,yshift=-\tcboxedtitleheight/2},
	boxed title style={size=small,colback=lightgray}
}{summary}

\newcommand\identity{1\kern-0.25em\text{l}}

\title{The Governance of Decentralized Autonomous Organizations: A Study of Contributors' Influence, Networks, and Shifts in Voting Power
}

\titlerunning{The Governance of Decentralized Autonomous Organizations} 

\author{	
Stefan Kitzler \inst{1,2}, 
Stefano Balietti \inst{3}, 
Pietro Saggese \inst{2,1}, 
Bernhard Haslhofer \inst{1} 
\and
Markus Strohmaier \inst{3,4,1} 
}

\institute{
	Complexity Science Hub Vienna \and
	AIT Austrian Institute of Technology \and
	University of Mannheim \and
    GESIS - Leibniz Institute for the Social Sciences
}

\authorrunning{S. Kitzler, S. Balietti, P. Saggese, B. Haslhofer, M. Strohmaier 
}

\begin{document}

\maketitle


\begin{abstract}

We present a study analyzing the voting behavior of contributors, or vested users, in Decentralized Autonomous Organizations (DAOs).
We evaluate their involvement in decision-making processes, discovering that in at least \SpacesNVPmoreFiftyPct/\% of all DAOs, contributors, on average, held the necessary majority to control governance decisions. 
Furthermore, contributors have singularly decided at least one proposal in \SpacesSelfAssociatesDecidedPct/\% of DAOs.
Notably, contributors tend to be centrally positioned within the DAO governance ecosystem, suggesting the presence of inner power circles.
Additionally, we observed a tendency for shifts in governance token ownership shortly before governance polls take place in \ShiftsProposalN/ (\ShiftsProposalPct/\%) of \ProposalsValidated/ evaluated proposals.
Our findings highlight the central role of contributors across a spectrum of DAOs, 
including Decentralized Finance protocols. Our research also offers important empirical insights pertinent to ongoing regulatory activities aimed at increasing transparency to DAO governance frameworks.

\end{abstract}


\section{Introduction}

DAOs represent organizational structures designed to offer an alternative, decentralized form of governance for decentralized applications (dApps) operating on Distributed Ledger Technologies (DLTs). The intention of DAOs is to circumvent central authorities and hierarchical structures that are prevalent in traditional organizations, and democratize the decision-making process by distributing voting rights through so-called governance tokens to community members~\cite{Chohan2017}.

Anecdotal evidence suggests that these intentions are not always met in practice. For instance, there are signs of a centralized power circle that has emerged within the Decentralized Exchange (DEX) service Sushiswap~\cite{sushi2021scandal}. Similarly, the governance of Arbitrum DAO proposed channeling tokens valued at 1 billion US dollars into their own treasury~\cite{unchained2023arbitrum}. The lending protocol Solend confiscated the funds of a prominent user who posed a risk to its financial stability~\cite{coin2022solana}. In another instance, major cryptoasset exchanges, significant entities in this context, reportedly colluded and leveraged investors' tokens to vote on the Steem platform~\cite{decrypt2020steem,messari2022theses}. Attempts at bribery have been noted among community members in governance forums~\cite{uni2021bribe}. Lastly, developers from the mixing service Tornado Cash are reportedly under investigation for financial crimes; it's alleged they manipulated its governance to circumvent the introduction of rigorous anti-money laundering controls~\cite{tornado2022arrest,tornado2022gov}.

It is well known that governance tokens are distributed primarily to team members, early investors, or protocol treasuries~\cite{barbereau2022defi}, and decision-making power can be concentrated in the hands of a few~\cite{Barbereau2023}. Earlier research has provided preliminary evidence on the involvement of DAO team members and developers in DAOs decision-making processes~\cite{Stroponiati2020governance,Jensen2021,Fritsch2022a}. However, there is a surprising gap in studies that systematically investigate the role of vested users in the governance of DAOs and how they determine their trajectories.

In this study, we focus on DAO \emph{contributors}, encompassing project owners, administrators, and developers. These contributors are involved in the technical realization of the dApp overseen by a DAO and thus can be viewed as vested users. Our aim is to empirically examine their influence in decision-making processes, the structure of their co-voting network, and any sudden shifts in majorities just before voting takes place. Our contributions and findings can be summarized as follows:

\begin{enumerate}
	
	\item We compiled a dataset comprising \VotersClean/ voters across \SpacesClean/ DAOs with \ContributionClean/ recognized contributions. Additionally, we cross-verified a subset of \VotesValidated/ votes from \ProposalsValidated/ proposals against their on-chain records, determining that \VotesConsistencyPct/\% of these were consistent.

	\item We introduce a metric to measure the \emph{involvement of contributors} in DAO voting:  
	in \SpacesNVPmoreFifty/ ($\,$\SpacesNVPmoreFiftyPct/\%) DAOs contributors held, on average, the necessary majority to steer governance. We also measured \emph{contributor self-decisions}, discovering that their votes were decisive in \SpacesSelfAssociatesDecided/ ($\,$\SpacesSelfAssociatesDecidedPct/\%) DAOs. 

	\item We analyze the co-voting structures of users through a network approach. Our findings indicate that contributors are more likely to be found towards the center of the DAO governance ecosystem. Furthermore, contributors are highly concentrated in a few communities formed by co-voting patterns.
  
	\item We observed \textit{majority shifts} in governance token ownership in \ShiftsProposalN/ (\ShiftsProposalPct/\%) out of \ProposalsValidated/ proposals in the days preceding the votes. The number of majority shifts increases sharply prior to governance polls, indicating last-minute token acquisitions.

\end{enumerate}


To the best of our knowledge, our study is the first to systematically investigate the role of contributors in the governance of DAOs. It underscores their pivotal role across various DAOs, including leading Decentralized Finance (DeFi) protocols. Beyond shedding light on centralization tendencies within DAO governance structures, our findings demonstrate that contributors possess the capability to effectively steer the direction of DAOs. These insights have significant implications regarding accountability. Moreover, they are relevant in the context of current regulatory initiatives aimed at pinpointing the individuals who either control or exert notable influence over DeFi operations or structures~\cite{IOSCO:2023a}.

We will release our dataset and the implementation of methodologies to ensure the reproducibility of our findings.


\section{Background, Definitions and Related Work}
\label{sec:background}

\subsection{Voting in Decentralized Autonomous Organizations}

Decentralized Autonomous Organizations (DAOs) are a novel form of governance model that has become popular in the crypto ecosystem since 2020. They can govern decentralized applications (dApps) and their associated smart contracts~\cite{zou2019smart}. 
Several Decentralized Finance (DeFi) protocols implement DAO governance models~\cite{Auer2023}, e.g., MakerDAO~\cite{maker2020whitepaper,Sun2022}, Uniswap~\cite{adams2021uniswap}, Sushiswap~\cite{sushiswap2022website}, and Compound~\cite{leshner2019compound}. 
DAOs can also operate without an underlying dApp~\cite{constitution2021documentation}.

DAO voting mechanisms can be divided into two primary categories: \emph{on-chain} and \emph{off-chain} voting. The former occurs directly on a DLT, through smart contracts implementing the voting logic. To vote, token holders delegate an address that can be controlled by another entity. This approach offers security and transparency, but transaction costs make it economically inefficient~\cite{Feichtinger2023,Dotan2023}. 
The latter takes place on centralized platforms like Snapshot~\cite{snapshot2023documentation}, and only the voting outcome is stored on the DLT. This method is more scalable, accessible and efficient, at the cost of higher centralization (e.g., concerns that the DAOs might not enforce the decisions, concurrent voting on different platforms, or non-tamper-proof databases).
Our study focuses on the Snapshot platform, the largest off-chain governance platform with a market share of over 90\% \cite{Wang2022b}.

Decision-making in DAOs is executed through voting on so-called \emph{improvement proposals} that can determine the evolution of the technical infrastructure~\cite{Uniswap2022deploy}, modify parameters affecting the economic incentives and design~\cite{curve2020inflation,Compound2023migration}, or reallocate funds managed by a DAO~\cite{Uniswap2023fees,Uniswap2023donation}. Governance users can participate in the voting by possessing specific tokens, known as \emph{governance tokens}, which represent their DAO membership and their proportional decision-making power. 

\subsection{Definitions}
\label{subsec:conceptualization}

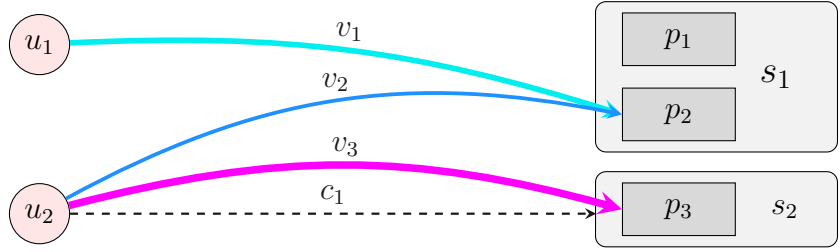
\begin{figure}[t]
	\centering
	\begin{tikzpicture}
	
	\tikzstyle{space} = [shape aspect=2,draw, rectangle, minimum height=1.cm, minimum width = 3cm, rounded corners, font=\large, align = right, anchor = base, text width=3cm]
	
	\tikzstyle{proposal} = [shape aspect=2,draw, rectangle, minimum height=0.7cm, minimum width = 1.5cm,font=\large]
	
	\tikzstyle{user} = [shape aspect=2,draw, circle, minimum height=0.8cm, minimum width = 0.8cm, fill = red!10,font=\large]
	
	\draw (9,-1) node[space, fill = gray!10, minimum height=2.cm] (S1) {\Large $s_1 \quad$};
	\draw (9,-2.75) node[space, fill = gray!10] (S2) {\large $s_2 \quad$};
	
	\draw (S1)+(-0.5,0.5) node[proposal, fill = gray!30] (P1) {\large $p_1$};
	\draw (S1)+(-0.5,-0.5) node[proposal, fill = gray!30] (P2) {\large $p_2$};		
	\draw (S2)+(-0.5,0) node[proposal, fill = gray!30] (P3) {\large $p_3$};	
	
	\draw (0,-0.5) node[user] (U2) {\large $u_1$};
	\draw (0,-2.75) node[user] (U3) {\large $u_2$};
	
	\path[-stealth,line width = .3mm, dashed, color = black!90] (U3) edge node[yshift = +0.25cm] {\large $c_1$} ++(7.4,0);
	
	\path[-stealth, line width = 0.75mm, color = {rgb,255:red,0;green,238;blue,238}] (U2) edge [bend left = +10] node[yshift = +0.25cm] {\large \textcolor{black!90}{$v_1$}} (P2.west);
	\path[-stealth, line width = 0.5mm, color = {rgb,255:red,30;green,144;blue,255}] (U3) edge [bend left = 20] node[yshift = +0.25cm] {\large \textcolor{black!90}{$v_2$}} (P2.west);
	\path[-stealth, line width = 1mm, color = {rgb,255:red,255;green,0;blue,255}] (U3) edge [bend left = 15] node[yshift = +0.25cm] {\large \textcolor{black!90}{$v_3$}} (P3.west);

	\end{tikzpicture}
	
	\caption{\textbf{Conceptualization of DAO voting.}
	A proposal $p$ introduces potential changes to a DAO space $s$, and users $u$ can 
	exert their decision-making power on them with their vote $v$
	(\protect\tikz[baseline=-0.5ex]\protect\draw[color = {rgb,255:red,0;green,238;blue,238}, -latex, line width = 0.5mm](0,0.05) -- (0.5,0.05);) and voting power $w$ indicated by the arrow thickness. Governance users can be vested by their contribution $c$ as \emph{owner}, \emph{administrator} or \emph{developer} to a space (\protect\tikz[baseline=-0.3ex]\protect\draw[black!90, -latex, line width = 0.5mm, dashed, thin](0,0.05) -- (0.5,0.05);).
	We denote their vested vote as $V^{P}_{SS}$ when they are contributors of the \emph{same-space}
	(\protect\tikz[baseline=-0.5ex]\protect\draw[color = {rgb,255:red,255;green,0;blue,255}, -latex, line width = 0.5mm](0,0.05) -- (0.5,0.05);)
	they are voting on,	and $V^{P}_{OS}$  when they are contributors of an \emph{other-space} (\protect\tikz[baseline=-0.5ex]\protect\draw[color = {rgb,255:red,30;green,144;blue,255}, -latex, line width = 0.5mm](0,0.05) -- (0.5,0.05);).
  }
	\label{fig:conceptual_model}
	
\end{figure}

We now present a conceptual model of DAO voting and introduce the key terminology and notation used throughout this paper, 
referring to DAOs as \emph{spaces}~\cite{snapshot2023documentation}. Figure~\ref{fig:conceptual_model} illustrates the entities: spaces, proposals and users; and it describes their relations of contribution and vote.

\begin{itemize}

	\item Let $\mathcal{U}$ be the set of all \textbf{users} exercising voting rights and $\mathcal{S}$ be the set of all \textbf{spaces}. Users can also be denoted as \textbf{voters} in this context.

	\item A \textbf{contribution} is a relation $\mathcal{C} \subseteq \mathcal{U} \times \mathcal{S} \times \mathbb{P} (\mathcal{T})$, where $(u,s,T) \in \mathcal{C}$, if a user $u$ contributes to a space $s$ in one or more role types $T \subseteq \mathcal{T} = \{
	 \text{owner},\text{administrator},\text{developer} \}$. Users can take multiple roles. A \textbf{contributor} is a user that has at least one contribution association to one space.

	\item A \textbf{proposal} is a relation $\mathcal{P} \subseteq \mathcal{S} \times \mathbb{P} (\mathcal{O}) \times \mathbb{P} ( \mathcal{F}) \times \mathbb{N^+}$, where $(s, O, F, h) \in \mathcal{P}$, if there is a proposed change to a space $s$ providing a set of choices or options $O$ to vote on, and a set of strategies $F$ to be applied for determining the outcome of a vote at a given block height $h$. The sets of options and strategies are defined as follows:
	
	\begin{itemize}

		\item $O^p \subseteq \mathbb{P} (\mathcal{O})$ denotes the set of possible options (choices) that can be selected during the voting phase on the improvement proposal $p$. In most cases, the alternatives are simply a yes/no answer (i.e., $O^p = \{Yes, No\}$).

		\item $F^p \subseteq \mathbb{P} (\mathcal{F})$ denotes the set of strategy functions that are applied to compute the voting power for the governance user issuing a vote.
	\end{itemize}

	\item A \textbf{vote} is a relation $\mathcal{V} \subseteq \mathcal{U} \times \mathcal{P} \times \mathcal{O} \times \mathbb{R}^{+}$, where $(u, p, o, m) \in \mathcal{V}$, if a user $u$ votes on a proposal $p$ by selecting an option $o \in O^p$, where $O^p$ denotes the set of options published as part of a specific proposal $p$. In rare cases, $o$ can become a vector, e.g., the associated voting strategies allow one to express multiple choices. Then, the magnitude vector $m$ characterizes the weighted preference of each option, and \(\sum_{i} m_{i} = 1\); if the vote expresses one single choice, $m$ is a scalar equal to 1. We further denote as $V^p \subseteq \mathcal{V}$ the set of all votes related to proposal $p$. We distinguish between two types of votes:

	\begin{enumerate} 

		\item A user can contribute and vote on an improvement proposal of the same space. We, therefore, denote $V^{P}_{SS} \subseteq V^p$ as the set of \textbf{same-space votes}, where, for all tuples $(u_i, p_i, o_i, m_i) \in V^{P}_{SS}$, a tuple $(u_j, s_j, T_j) \in \mathcal{C}$ such that $u_i = u_j$ and $s_n = s_j$ for $p_i^{s_n}$ exists, i.e., the users $u_i$ equals $u_j$ and $s_n$ of proposal $p_i$ equals the space $s_j$.

		\item A user can also contribute to one space and vote on an improvement proposal for another space. We denote $V^{P}_{OS} \subseteq V^P$ as the set of \textbf{other-space votes}, where for all tuples $(u_i, p_i, o_i, m_i) \in V^{P}_{OS}$, a tuple $(u_j, s_j, T_j) \in \mathcal{C}$ such that $u_i = u_j$ exists, but there does not exist one where additionally $s_n = s_j$ is fulfilled for $p_i^{s_n}$. Note that: $V^{P}_{SS} \cap V^{P}_{OS} = \emptyset$.
		
		\item Finally, we denote $V^{P}_{C} = V^{P}_{SS} \cup V^{P}_{OS}$ as the set of contributor votes.

	\end{enumerate}

	\item The \textbf{voting power} is the weight $w$ assigned to an option $o$ and characterizes the influence of a vote $v$. It is determined by the strategy function $f : \mathcal{V} \times \mathbb{N^+} \rightarrow \mathbb{R^+}$ of the vote $v$ at block height $h$. For proposals with multiple functions, the weight is defined by their sum $F^p(v,h) := \sum_{f \in F^p} f(v,h)$.

	\item Finally, the options $O^p$ can be ranked by aggregated voting power $w$. We denote as the \textit{outcome} the options $\hat{O}^p = [ \hat{o}_1^p, \, \hat{o}_2^p, \, \dots ]$ ranked in descending order by voting power, and denote $\hat{o}_1^p$ as the \textit{decision}, i.e., the option having the highest accumulated voting power for the proposal $p$.
	
\end{itemize}

\subsection{Related work}

Prior research has extensively documented that the ownership of governance tokens is highly concentrated~\cite{Stroponiati2020governance,Jensen2021,Nadler2020,barbereau2022defi,Dotan2023}, as a result of intentional design decisions and market dynamics (governance tokens carry a market price and can be traded). Furthermore, their total supply, the monetary policy, and the initial token allocation affects their distribution; finally, mechanisms such as airdrops~\cite{frowis2019operational} further favor early participants and DAO members.

Studies focusing specifically on on-chain DAO voting confirm that governance tokens are highly concentrated. Furthermore, they show that users rarely exercise voting rights \cite{Barbereau2023}, and that individuals who possess the potential power to alter outcomes rarely exercise it \cite{Fritsch2022a,Feichtinger2023}. %
Two related works identify the existence of voters' coalitions in MakerDAO \cite{Sun2022,Sun2023}.
A preliminary study reports examples of voters who held governance tokens for the duration of a single proposal life-cycle~\cite{Dotan2023}.

Our work is closely related to studies on off-chain voting. Wang et al.\cite{Wang2022b} delivered a comprehensive overview of the voting platform Snapshot. Laturnus \cite{Laturnus2023} utilized data from Snapshot and DeepDAO to investigate the economic performance of DAOs in relation to ownership concentration and voting participation.


While earlier research has provided preliminary evidence on the involvement of DAO team members and developers in DAOs decision-making processes, none of these studies has systematically investigated the role of vested users in DAOs and their influence in determining their trajectories. This knowledge gap serves as the motivation for our study.


\section{Data}\label{sec:data_collection}

To analyze the involvement of contributors in DAO decision-making through voting, we gather data from the following sources:
Snapshot, Ethereum blockchain, Ethereum Name Service (ENS) and The Graph. We combine them to identify contributions, as defined in Section~\ref{subsec:conceptualization}. Then, we clean, verify, and validate our dataset, as summarized in Table~\ref{tab:datasets}. Additional details on the entire data preparation process and contribution identification are reported in Annex~\ref{sec:Annex_Data}.

\begin{table}[t]
	\centering
	\begin{tabular*}{\columnwidth}{@{\extracolsep{\fill}}lrrr}
\toprule
{ } & {Raw} & {Cleaned} & {Validated} \\
& & (Sections 4 \& 5) & (Section 6) \\
\midrule
Spaces $ \mathcal{S}$ & \num{12294} & \num{872} & \num{357} \\
Voters  $\mathcal{U}$ & \num{1603994} & \num{986557} & \num{119413} \\
Contributions $\mathcal{C}$ & \num{11949} & \num{7478} & \num{3927} \\
Proposals $ \mathcal{P}$ & \num{76851} & \num{35124} & \num{8116} \\
Votes $ \mathcal{V}$ & \num{8365707} & \num{5240622} & \num{438668} \\
Contributor votes $ \mathcal{V_C}$ & \num{316900} & \num{191507} & \num{22878} \\
\bottomrule
\end{tabular*}

	\vspace{0.2cm}
	\caption{\textbf{Dataset summary.} The \textit{raw} dataset combines Snapshot data on voters $\mathcal{U}$ with additional sources to identify the contributions $\mathcal{C}$ and quantify their voting activity $ \mathcal{V_C}$. Users vote ($ \mathcal{V}$) on improvement proposals $ \mathcal{P}$ to DAO spaces $ \mathcal{S}$. To focus on DAOs with mature governance structures, we \emph{cleaned} and \emph{validated} the data set using selected proposals with Ethereum on-chain data.}
	\label{tab:datasets}
\end{table}

\paragraph{Raw dataset.}\label{sec:raw}

We obtained \VotersRaw/ \emph{DAO voters} and their wallet addresses from the Snapshot dataset. These have cast \VotesRaw/ votes from Nov-2020 to Dec-2022 on \ProposalsRaw/ \emph{proposals}, using \StrategiesN/ distinct voting strategies in \SpacesRaw/ \emph{DAO spaces}.
Next, we identify voters' contributions to DAOs by joining their addresses with additional data. We extract their respective roles $\mathcal{T}$ by retrieving the domain \textit{owner} address from ENS references, the \textit{administrators'} addresses from Snapshot, and the creators, or \textit{developers}, of code accounts (CA) from the blockchain transaction for all space-related CA from Snapshot.

\paragraph{Cleaned dataset.}\label{sec:cleanedData}

We found that \SpacesSingleProposalPct/\% of DAO spaces have one proposal only, \SpacesFollowersLessFivePct/\% have less than five followers and 
\SpacesLessTwoVotesPct/\% have at most two voters. 
We consider these as indicators for immature governance structures. Therefore, we apply a cleaning procedure by incorporating related benchmarks that assess minimum requirements to include mature DAOs in the data set. We also remove non-final proposals and restrict to proposals using the \textit{single-choice} voting.

\paragraph{Validated dataset.}\label{sec:validation}

Previous studies have shown inconsistencies between the reported and actual on-chain data, including instances of flawed data records within a Blockchain explorer~\cite{He2023}, emphasizing the need for a validation framework. Therefore, we validate the consistency between the voting power values computed by Snapshot and the ground truth reflected in on-chain data.

We focus on the Ethereum Blockchain, the most relevant one for Snapshot~\cite{Wang2022b} in terms of expressed voting power, and only consider proposals that are almost entirely\footnote{
The selected strategies cover more than 99\% of the voting power in the proposals.
} covered by strategies bound to $F' \subseteq \mathbb{P}( \{f^{erc20}, f^{erc721}, f^{eth} \} )$. With this approach, we could verify that \VotesConsistency/ (\VotesConsistencyPct/\%) of \VotesToValidate/
Ethereum Snapshot weights are correct.


\section{Influence of contributors on DAO governance}
\label{sec:influence}

\subsection{Contributor involvement}\label{sec:involement}

Contributors are vested users, having intuitively higher incentives to be involved in the decision-making in DAOs. 
We analyze their involvement by measuring their voting power exercised in proposals. 
As discussed in Section \ref{subsec:conceptualization}, the voting power $w_i$ is the weight assigned to an option $o_i$ and characterizes the influence of a vote $v_i$. In most cases, it is equal to the amount of governance tokens held by the user. Recall that the weight $w_i$ is the result of a proposal strategy function $F^p(v_i, h)$ applied on a vote $v_i$ at a specific block height $h$.

We compute the involvement of contributors in a given space by averaging the share of voting power they have in proposals associated with that space. Since minted amounts of governance tokens, their prices, and their distributions across user vary, we normalize voting power by total voting power across proposals as follows: $\tilde{w_i} =  w_i \times (\sum_{v_l \in V^p} w_l)^{-1}$. Next, we consider the set of contributors $V^p_{C}$, that includes \emph{same-spaces} voters $V^{p}_{SS}$ as well as \emph{other-space} $V^{p}_{OS}$. 
Then, we compute the fraction of weights controlled by contributors $\tilde{w}^{p}_C$ (\ref{eq:relVpContr}) and finally obtain the \textbf{contributor involvement} $\bar{w}^{s}_C$ (\ref{eq:avgContrInv}) as the average weights of contributors' votes for all proposals $P$ in a DAO space $s$.

\vspace{0.25cm}

\noindent\begin{minipage}{.5\linewidth}
	\begin{equation}
	\tilde{w}^{p}_C = \sum_{v_i \in V^p_{C}} \tilde{w_i}, \label{eq:relVpContr}
	\end{equation}
\end{minipage}
\begin{minipage}{.5\linewidth}
	\begin{equation}
	\bar{w}^{s}_C = \, |P|^{-1} \sum_{p \in P} \tilde{w}^{p}_C. \label{eq:avgContrInv}
	\end{equation}
\end{minipage}

\vspace{0.25cm}

%
To give an example, let's assume proposal $p$ has four votes $\{v_1, v_2, v_3, v_4\}$ with normalized voting powers $\{.1, .4, .3, .2\}$, where $\sum\tilde{w_i} = 1$. Supposing that the first two voters are contributors, ie $V^p_{C} = \{v_1, v_2\}$, then $\tilde{w}^{p}_C = .1 + .4 = .5$.

\begin{figure}[t]
	\includegraphics[scale=0.99, width=\columnwidth]{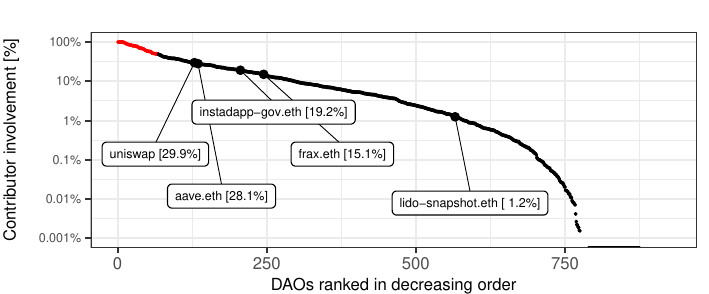}
	\caption{\textbf{Contributor involvement across DAO spaces}. The DAOs are ranked by contributor involvement $\bar{w}^{s}_C$ (\protect\tikz[baseline=-0.5ex]\protect\filldraw[black] (0,0) circle (1.5pt);) from highest (left) to lowest (right). Some high-TVL dApps (\protect\tikz[baseline=-0.5ex]\protect\filldraw[black] (0,0) circle (3pt);) are annotated for illustrative purposes and contributor involvement of more than \num{50}\% is colored (\protect\tikz[baseline=-0.5ex]\protect\filldraw[red] (0,0) circle (1.5pt);).
	}
	\label{DAO_assoc_mean_vp}
\end{figure}

We determine $\bar{w}^{s}_C$ for all spaces on the cleaned data set, and show the results in Figure~\ref{DAO_assoc_mean_vp}. The DAO spaces are ranked by contributor involvement in descending order. Thus, the involvement of contributors is high for the DAOs on the left-hand side and low for the DAOs on the right-hand side. For illustration purposes, we highlight the data points representing top DeFi protocols in terms of Total Value Locked (TVL), such as Aave, Uniswap or Instadapp. 

Our results show that the involvement of contributors in terms of average voting power is relatively low for most DAOs. The median value is \VpMedian/\% and the standard deviation is \VpStd/. However, for \SpacesNVPmoreTen/ spaces, the relative voting power of contributors is higher than 10\% and for \SpacesNVPmoreFifty/ DAOs it is higher than \num{50}\%. In these spaces, the contributors have, on average, a majority of voting power and can determine single-handedly the outcome of proposals. In \SpacesNVPmoreHundred/ spaces, the contributors were the only voters with \num{100}\% voting power.

\subsection{Contributor self-decisions}
\label{sec:selfvoting}

Knowing that DAO contributors are involved in decision-making, we now investigate to what extent they decide on proposals related to their own spaces. Thus, we concentrate our analysis on the votes cast by users that contributed on improvement proposals of the same spaces ($V^{p}_{SS}$, see Section~\ref{subsec:conceptualization}), which we herein denote as ``self-votes''. Furthermore, we also consider the choices they made with their votes and their influence on the outcome of a proposal.

Recall that the \emph{decision} of a proposal poll is determined by the option with the highest voting power $\hat{o}_1^p$ within the ranked outcome $\hat{O}^p = [ \hat{o}_1^p, \, \hat{o}_2^p, \, \dots ]$. We denote the set of \emph{decisive self-votes} as $V^{p}_{D}$, where the option $o$ of $V^{p}_{D}$ is the winning choice $\hat{o}_1^p$. 

We are specifically interested in the self-votes where the decision-making was dominated by contributors, which we denote as \emph{contributor self-decisions}. Intuitively, for a given space $s$, we determine the share of selected proposals based on two joint conditions. First, we consider the weight of contributor votes within a decision and select those proposals where contributors have a relative majority ($\geq$ 50\%). Second, we consider also the second-ranked option and select those proposals where the weight of contributors in the decision is higher than the weight of the second ranked option. The underlying intuition is that in a head-to-head race between options, contributors might want to outweigh and overrule a leading option. 

More formally, we define the set of \textit{decisive self-votes} $V^{p}_{D} = V^p_{\hat{o}1} \cap V^p_{SS}$ and also the complement set $V^p_{CV} = V^p_{\hat{o}1} \, \backslash \, V^{p}_{D}$; we identify the fractions of relative voting power for decisive self-votes (\ref{eq:relVpSV}), for the complement set (\ref{eq:relVpCV}) and for the second choice $\hat{o}_2^p$ (\ref{eq:relVpO2}) as

\noindent\begin{minipage}{.3\linewidth}
	\begin{equation}
	\tilde{w}^{p}_{D} = \sum_{v_i \in V^p_{D}} \tilde{w_i}, \label{eq:relVpSV}
	\end{equation}
\end{minipage}
\begin{minipage}{.3\linewidth}
	\begin{equation}
	\tilde{w}^{p}_{CV} = \sum_{v_i \in V^p_{CV}} \tilde{w_i}, \label{eq:relVpCV}
	\end{equation}
\end{minipage}
\begin{minipage}{.3\linewidth}
	\begin{equation}
	\tilde{w}^{p}_{\hat{o}2} = \sum_{v_i \in V^p_{\hat{o}2}} \tilde{w_i}. \label{eq:relVpO2}
	\end{equation}
\end{minipage}	

\vspace{0.25cm}

For a given space $s$, we can define the \emph{contributor self-decisions} $\delta^s$ as follows:

\begin{align}
\delta^s := 
|P|^{-1} 
\,
\sum_{p \in P} [ \,
(\tilde{w}^{p}_{D}  > \tilde{w}^{p}_{CV} )
\land
(\tilde{w}^{p}_{D} > \tilde{w}^{p}_{\hat{o}2})
\,
]
\quad.
\label{eq:deltaS}
\end{align}

\begin{figure}[t]
	\includegraphics[width=\columnwidth]{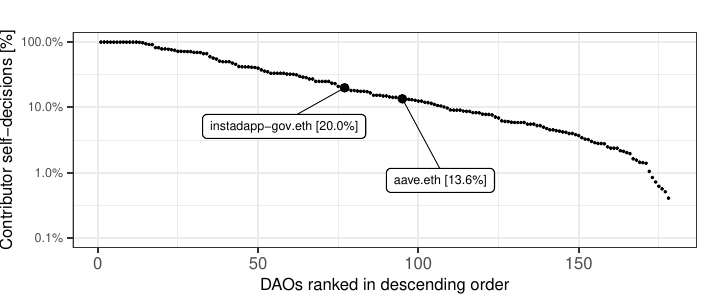}
	\caption{\textbf{Contributor self-decisions across DAO spaces.} The \SpacesSelfAssociatesDecided/ DAOs are ranked by contributor self-decisions $\delta^s$
	(\protect\tikz[baseline=-0.5ex]\protect\filldraw[black] (0,0) circle (1.5pt);)	
	in descending order, with a threshold of \num{0.1}\%. The y-axis represents the fraction of proposals in which DAO contributors voted and decided their outcome with dominant voting power. Some high-TVL dApps 
	(\protect\tikz[baseline=-0.5ex]\protect\filldraw[black] (0,0) circle (3pt);)
	are annotated for illustrative purposes.}
	\label{fig:DAO_dec_self_maj}
\end{figure}

Figure \ref{fig:DAO_dec_self_maj} shows the results with DAOs ranked by self-decisions in descending order. Note that we introduced thresholds and only show spaces with self-decisions above \num{0.1}\%. This gives us \SpacesSelfAssociatesDecided/ (\SpacesSelfAssociatesDecidedPct/\%) different spaces where contributors of the same DAO decided on at least one proposal on their own. 
In total \ProposalsSelfAssociatesDecided/ out of \ProposalsClean/ proposals were decided by governance users who contributed and voted on the same DAO.
Annex \ref{sec:Annex_Influence} provides more details and analyses on involvement and self-decisions.

\section{Co-voting networks}
\label{sec:network}

In Section~\ref{sec:influence}, we measured the involvement of  contributors in decision-making for each space separately; however, users can contribute to multiple proposals and DAO spaces.
Herein, we conduct a network-based analysis of users' co-voting patterns across DAOs.
We analyze topological features such as centrality measures and community structures that may indicate whether contributors occupy a central role in the DAO voting ecosystem.

\paragraph{Networks construction.}

The basis for the investigation is the bipartite network $G_{PU}$ that links users $\mathcal{U}$ to the proposals $\mathcal{P}$ they voted on, having options $\mathcal{O}$ as edge features.
We derive \textit{co-voting networks} as a monopartite projection of the $G_{PU}$ on voters, by creating a network of users with weighted links that represent the number of proposals they voted together. 
We introduce a global threshold $T$ on links to focus on users that systematically voted together on the same proposals and for computational reasons. 

Ultimately, we build four co-voting networks crossing DAOs and votes as shown at the top of Table~\ref{table_network_descriptive}, namely:

\begin{itemize}
	\item \textbf{$\boldsymbol{G_{AA}}$} is the entire co-voting network, containing all votes, regardless of users' choices, and all spaces;
	\item \textbf{$\boldsymbol{G_{AW}}$} is the co-voting network of decision-makers. It only takes into account votes $v_i$ for the \textit{winning} decision $\hat{o}_1^p$ (we hypothesize that co-voting patterns may be especially relevant among users who voted for the winning outcome); 
	\item \textbf{$\boldsymbol{G_{TA}}$} is the co-voting network of all votes of the top-100 DAOs by TVL.
	\item $\boldsymbol{G_{TW}}$ is the co-voting network of decision-makers in the top-100 DAOs by TVL, was constructed in the same fashion of $G_{AW}$.
\end{itemize}

\paragraph{Network descriptive statistics.}

We utilize the cleaned data set of \VotesClean/ voting relations on \SpacesClean/ DAOs and \ProposalsClean/ proposals. We create the networks using the threshold $T = 10$ on the links among voters; despite this, 
given the inherent computational challenge on computing metrics on the large network $G_{AA}$, we focus mainly on the remaining three networks.

In all networks, we identify small-world features, i.e., they are characterized by the presence of several hubs conveying information rapidly across connected communities \cite{watts1998collective}.  Interestingly, the share of contributor nodes and edges increases in the Top-100 networks (see bottom rows in Table~\ref{table_network_descriptive}), suggesting both that they are more active and that they tend to have a larger weight in shaping the outcomes of proposals of the most important DAOs, rather than the peripheral ones. The rest of the section tries to confirm whether this intuition is true. 
Appendix~\ref{appendix_network} contains additional network statistics and analyses.

\begin{table}[t]\centering
	\centering

\begin{tabular*}{\columnwidth}{@{\extracolsep{\fill}}lrrrr}
    \toprule
    Network &  $G_{AA}$ &  $G_{AW}$ & $G_{TA}$ &  $G_{TW}$\\
    \hline
        Daos &
        All & All & Top-100 & Top-100 \\
                Votes &  All &  Winning & All & Winning \\
    \midrule
    \midrule
            Num Nodes & \num{104863} & \num{75879} & \num{20401} & \num{14494} \\
            Num Edges & \num{739813062} & \num{107374710} & \num{19917792} & \num{6045065} \\
        Avg. Degree & \num{14110.09} & \num{2830.16} &  \num{1952.63} & \num{834.15} \\
        \midrule
        Contr. Nodes &  1.29\% & 1.45\% &  3.25\% & 4.5\% \\
        Contr. Edges & 1.61\% & 1.76\% &  3.4\% & 8.0\% \\
        \bottomrule
\end{tabular*}

	\vspace{0.2cm}
	\caption{\textbf{Network statistics of four co-voting networks.} The top of the table defines the four networks as a unique combination of two features: DAOs and Votes. \textit{Top-100} DAOs are ranked by total value locked (TVL); \textit{Winning} votes are votes for the choice that ultimately won the majority of voting power.}
	\label{table_network_descriptive}
\end{table}

\subsection{Centrality of contributors}\label{sec_networks_centrality}

\begin{figure}[t]
	\centering
	\includegraphics[width=0.95\columnwidth]{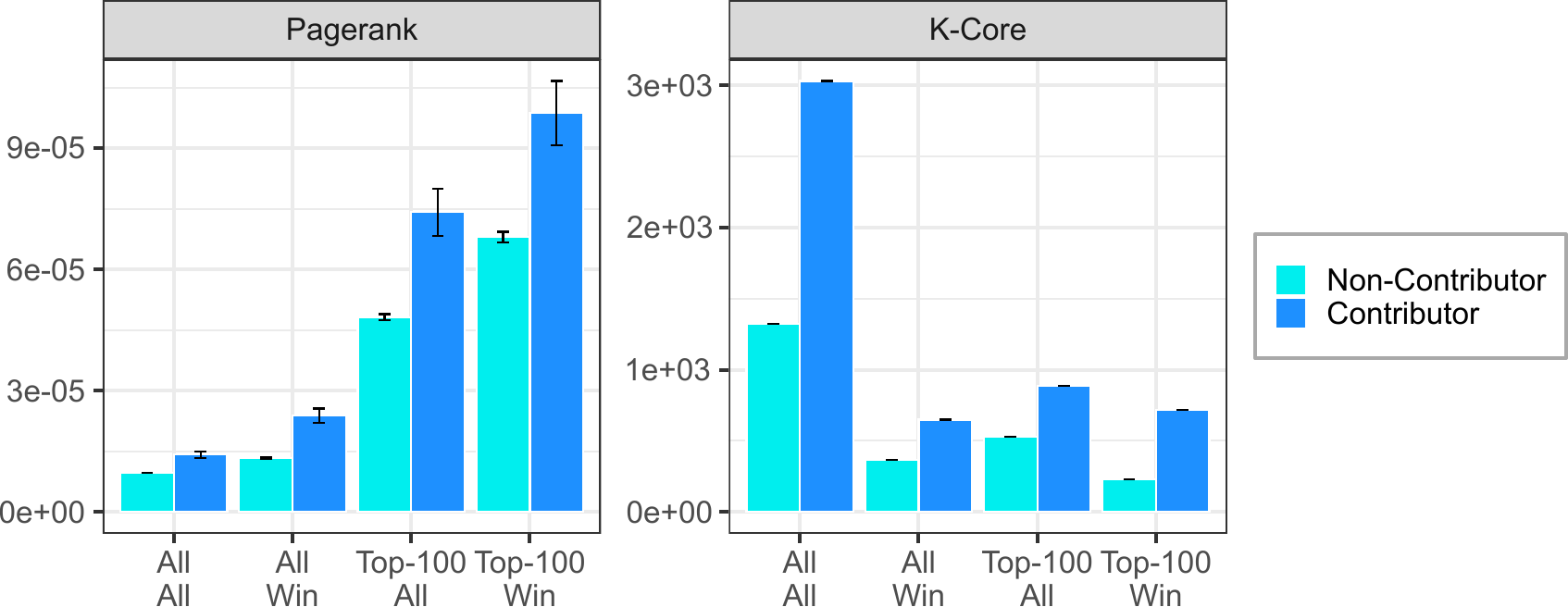}
	
	\caption{\textbf{Pagerank and k-core statistics in the four co-voting networks}. Contributors tend to have higher pagerank and k-core across networks. All centrality measures make use of edge weights and are applied to the giant component; k-core statistics use geometric mean to limit the effect of outliers. Error bars are 95\% confidence intervals of the means.}
	\label{fig_centrality_all} 
\end{figure}

To understand the influence of contributors on governance voting, we computed 
several network centrality measures for contributor and non-contributor nodes, namely pagerank, closeness, eigenvector, and betweenness centrality, 
and a k-core analysis. We present pagerank in Figure~\ref{fig_centrality_all} as well as the k-core.
Across all four networks, contributor nodes score higher in centrality in all measures but eigenvector for $G_{AW}$. These differences are generally highly significant (t-test $p< 0.001 $), only the betweenness centrality shows more variability ($p< 0.05$ for $G_{TA}$, and $p< 0.1 $ for $G_{TW}$).

We also computed the k-coreness of contributors. A high k-core indicates direct connections with other nodes with high k-core nodes, that is, nodes with at least degree $k$. Across all networks, contributors have, on average, significantly higher k-core (t-test $p < 0.001$). For these statistics we chose to use the geometric means because they are less sensitive to outliers. In fact, contributors are generally less frequent in the portion of the distribution with the lowest k-core, however, there exists a few clusters of mainly non-contributors with very high k-core, which would skew the results.


\subsection{Communities of contributors}\label{sec_neworks_communities}

\begin{figure}[t]
	\centering
	\includegraphics[width=0.99\columnwidth]{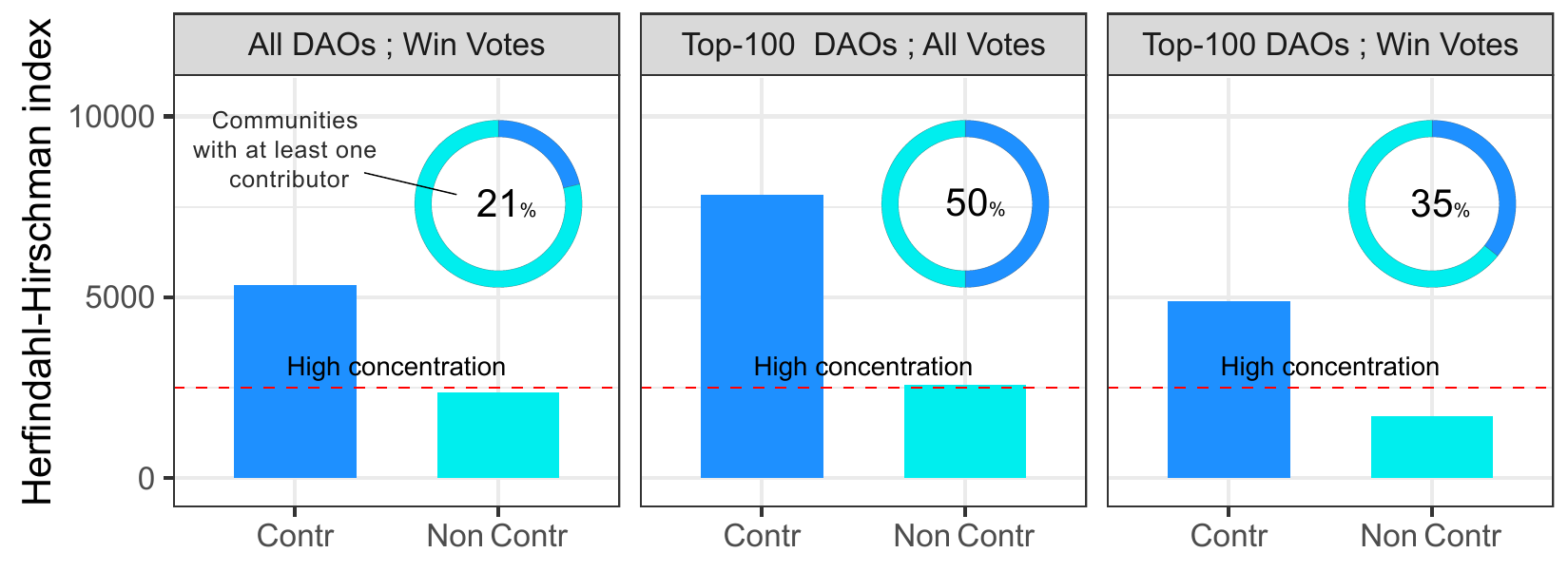}
	\caption{\textbf{Concentration of contributors across network communities.} The bar plots show the Herfindahl-Hirschman concentration index for the distribution of contributors 
	(\protect\tikz[baseline=0ex]\protect\draw[fill={rgb,255:red,30;green,144;blue,255},line width=1pt]  (0,0) rectangle ++(0.2,0.2);) 
	and non-contributors (\protect\tikz[baseline=0ex]\protect\draw[fill={rgb,255:red,0;green,238;blue,238},line width=1pt]  (0,0) rectangle ++(0.2,0.2);) 
	to communities 
	assigned by the Louvain community detection algorithm. The inset donut plots show the share of communities with at least one contributor; in all networks, contributors are concentrated in a few of them.}
	\label{fig_louvain_concentration}
\end{figure}

\begin{figure}[t]
	\centering
	\includegraphics[width=0.7\linewidth]{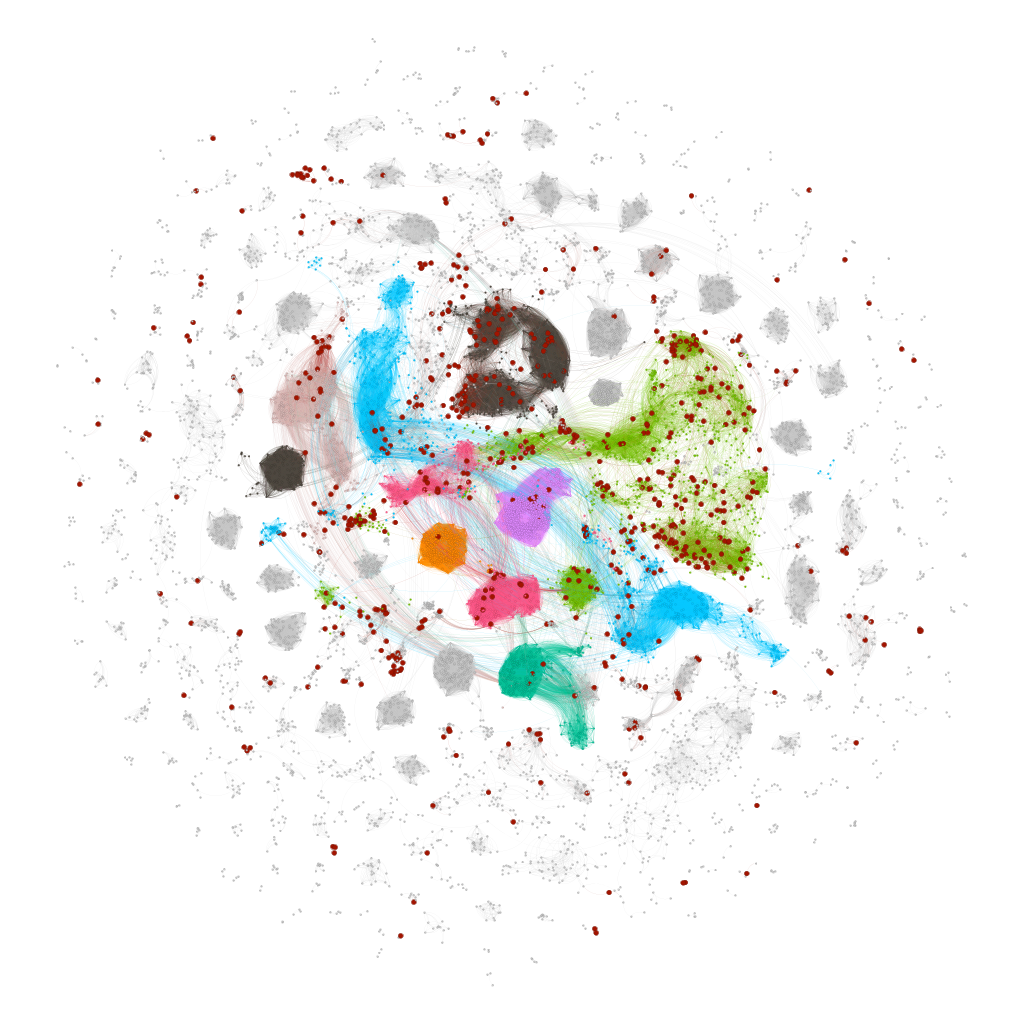}
	\caption{\textbf{The co-voting network of the Top-100 DAOs by TVL (winning votes only).} The colors
	identify the largest communities obtained by optimizing modularity, while smaller communities are in light gray; contributor nodes are colored in dark 
	red (\protect\tikz[baseline=-0.5ex]\protect\filldraw[contributor_red] (0,0) circle (1.5pt);). Contributors are not uniformly distributed across communities, but they tend to cluster in a few of them in the middle of the graph, suggesting a higher network centrality. Network plotted using the
	OpenOrd layout algorithm in Gephi \cite{bastian2009gephi}, after removing redundant edges 
    following \cite{nocaj2015untangling}.
}
	\label{fig_network_top100win}
\end{figure}

To understand the presence of hidden co-voting formations, we performed the Louvain community detection method on three co-voting networks. This method 
optimizes the modularity of the graph so that the connections within each community are dense, while the connections across communities are sparse. As a result, each node in the graph is uniquely assigned to a community. We then tested if contributors can be found with equal probability in all communities or whether they are more likely to cluster together in a few of them. To answer this question, we computed the Herfindahl-Hirschman index of market concentration on the distribution of contributors to communities. A higher value of this index indicates a more concentrated market, that is a distribution with fewer groups or communities dominating in size. Figure \ref{fig_louvain_concentration} indeed shows a very high concentration level for contributors in all networks with a peak
above \num{7000} (below \num{1500} is considered well-mixed, between \num{1500} to \num{2500} is
moderately concentrated, and above \num{2500} is highly concentrated). Counting the communities with at least one contributor (the donut plot inside each panel of Figure. \ref{fig_louvain_concentration}), contributors
are to be found only in about 21-50\% of all detected communities. A Pearson's Chi-squared
test indicates a significant deviation from chance in all networks ($p < 0.001$,
 with \num{100000} bootstrapped replicates). 
Figure \ref{fig_network_top100win} visually confirms this result for the $G_{TA}$ network: contributors (dark red) tend to cluster in a few central communities.

\section{Pre-voting power shifts}
\label{sec:shifts}

Governance tokens are cryptoassets and, consequently, can be purchased and sold. Furthermore, previous works provide preliminary evidence that users may hold their voting rights only for the duration of single proposals~\cite{Dotan2023}.
Therefore, we hypothesize that changes in the ownership distribution shortly before the voting power is determined could indicate attempts to acquire additional power to influence a proposal's decision. We investigate to what extent users, and especially contributors, acquire voting rights shortly before the poll execution. 

For proposals that rely on on-chain data, it is possible to access the current and historical token balances of voters. Thus, we determine their voting power at earlier points in time by re-implementing the proposal strategies on historical data and comparing it to their actual voting power. 
Note that we recompute it considering that the voter $v_i$ still selects the same option $o_i$. 
Assuming that users' holdings do not fluctuate with high frequency, we sample their token balances on a daily basis\footnote{Days are approximated assuming that blocks are mined on average every 15 seconds, i.e., $1 \text{d} \doteq (86400 / 15) \, \text{blocks}$.}. More formally, given a proposal $p$, for each voter $u_i$ we denote the actual voting power $w_{i}(h_\tau)$ as the voting power at the block $h_\tau$ of the vote execution, and $\hat{O}^p(h_\tau) = [ \hat{o}_1^p(h_\tau), \, \dots ]$ as the actual ranked outcome.
Next, for each of the 100 days preceding the vote on the proposal $p$, we recompute the users' historical voting power $w_{i}(h_{\tau-t})$ and the resulting hypothetical ranked outcome $\hat{O}^p(h_{\tau-t})$, where $h_{\tau-t}$ is a block representative of the $t^{th}$ day before the poll. We thus compare $\hat{O}^p(h_{\tau-t})$ to $\hat{O}^p(h_{\tau-t-1})$ and determine whether there was a \textit{majority shift} if $\hat{o}_1^p(h_{\tau-t}) \neq \hat{o}_1^p(h_{\tau-t-1})$. 
Finally, we measure the number of majority shifts across proposals.
Since we are correlating against on-chain data, we utilize the validated dataset described in Section~\ref{sec:data_collection}, covering \ProposalsValidated/ (\ProposalsValidatedPct/\%) proposals. We therefore emphasize that the findings reported in this Section are a lower boundary estimation.

\begin{figure}[t]
	\includegraphics[width=0.95\columnwidth]{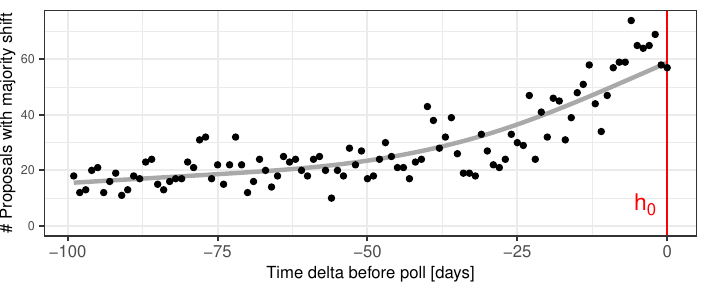}
	\caption{\textbf{Majority shifts occur in temporal proximity of polls}. Majority shifts occur when the voters, shortly before a proposal, trade enough governance tokens to swing the final outcome of the poll. We focus on the \textit{validated dataset} and identify majority shifts up to 100 days before the votes. In temporal proximity to the proposals, the number of shifts increases, indicating \textit{last-minute} voting power acquisition.}
	\label{NShiftsBeforeSnapshot}
\end{figure}

In total, we found majority shifts for \ShiftsProposalN/ (\ShiftsProposalPct/\%) proposals in \ShiftsSpaceN/ DAOs in the 100 days before the poll. The median number of shifts per proposal is 1, 
with a standard deviation of \ShiftsProposalStd/, and the maximum number of shifts for a single proposal is 30. 
To investigate whether the majority shifts are more frequent in the proximity of vote executions, Figure~\ref{NShiftsBeforeSnapshot} reports the aggregated count of majority shifts across proposals as a function of the time distance from the vote execution. We observe a constant or slightly increasing trend in farther dates from $-100$d to $-50$d, and a clearly increasing trend the closer time gets to the vote date $0$d. 
This indicates that the trading of governance tokens increases shortly before polls and that users might trade voting power to decide the outcome of the proposal in their preferred way. We acknowledge, however that we only identify a pattern and further research is required to better investigate this phenomenon.

Finally, we examine the participation of contributors in the proposals with majority shifts. Out of \ShiftsProposalN/ proposals with majority shifts, \ShiftsContributorN/ contributors associated with \ShiftsDAOContributorN/ different DAO spaces voted in \ShiftsContributorsProposalN/ (\ShiftsContributorsProposalPct/\%) proposals. 


\section{Discussion and Conclusions}

Our study augments the existing body of knowledge on decision-making in DAOs. It substantiates the results of previous studies, highlighting that the distribution of governance tokens is highly concentrated~\cite{barbereau2022defi,Fritsch2022a}, and the exercise of voting rights is very low~\cite{Barbereau2023}. Going beyond these findings, we discovered evidence that contributors, who are essentially users vested in DAOs, are involved in decision-making and, in some cases, have the power to effectively influence the trajectories of DAOs. Furthermore, we provide evidence that contributors are more likely found at the center of the DAO governance ecosystem and that majority shifts happen, especially shortly before the votes.

These findings have several implications. First, they suggest that contributors are overrepresented in the decision-making process of certain DAOs compared to other governance users. This is in line with known concerns that contributors may have differing interests and that users with smaller stakes might be discouraged from voting~\cite{buterin2021coinvoting}. Second, we found only limited evidence for DAO contributors influencing other, possibly competing, spaces. This is relevant because a rational governance token holder vested in a space might be interested in voting against proposals benefiting the evolution or adoption of other, competing DAOs (see~\cite{ford2019rationality}). Third, we found evidence of co-voting patterns among contributors, which is an indicator of the existence of inner circles of power in DAOs. 
These findings refute the conventional wisdom that DAOs are decentralized and run autonomously without being under anyone's control. This is relevant for resolving questions of accountability, as vested users and large governance token holders may be considered members of a legally recognized entity and therefore responsible for the underlying dApp. The ongoing Tornado Cash investigation claiming that the developers influenced its governance is a prime example of this line of argumentation.

Our work clearly faces some limitations and opens directions for further research. Currently, it focuses on off-chain voting and on one platform alone (Snapshot), whilst voting is executed also on-chain on multiple DLTs, as well as on other off-chain platforms. Extending the study to other governance platforms and to on-chain DLT voting would be a straightforward improvement. Furthermore, our results provide preliminary evidence that majority shifts take place before voting. It would be important to investigate and explain more formally the factors influencing governance participation and voting, also combining on-chain data with traditional methods surveying crypto users \cite{state_of_crypto_22_report_2023,auer2022distrust}. Lastly, we view the contributors in our dataset as a lower boundary. Incorporating more data sources, such as Github, would likely elevate this baseline.

Who governs DAOs? This question has been the driving force behind our research and is also a significant concern for regulators currently formulating policy recommendations for Decentralized Finance (DeFi) systems~\cite{IOSCO:2023a}. Although our study does not aim to unveil the identities of the responsible individuals, as regulatory efforts suggest, it offers a systematic investigation into the role of vested users in the governance of DAOs. This, in turn, can provide valuable insights and inform ongoing regulatory debates on that topic.


\clearpage

\bibliography{literature.bib}

\clearpage

\appendix


\section{Data}\label{sec:Annex_Data}

\setcounter{figure}{0}
\renewcommand\thefigure{\thesection.\arabic{figure}}    
\setcounter{table}{0}
\renewcommand\thetable{\thesection.\arabic{table}}    

This Annex provides additional material to Section~\ref{sec:data_collection}. We start by listing more detailed information on the data sources (\ref{sec:Annex_Sources}) and how we extracted the contributor roles (\ref{sec:Annex_Contributors}). Additionally, we include how we gathered information on Total Value Locked (TVL) of DAOs and describe the scrape and join procedure in (\ref{sec:tvl}). 
We describe the cleaning procedure of the \emph{raw} data in (\ref{sec:Annex_Cleaning}), add more information on the voting strategies (\ref{sec:strategies}) and finally outline the data validation procedure (\ref{sec:tokens}).

\subsection{Sources}\label{sec:Annex_Sources}

We gathered data from the following sources:

\begin{itemize}
	
	\item \textbf{Snapshot}: We received a comprehensive database dump from the Snapshot off-chain voting platform, provided by the platform's developers. It encompasses all information about the DAO spaces, associated proposals, and votes on each proposal, spanning from the platform's inception in November 2020 up until the 7$^{th}$ of December 2022.
	
	\item \textbf{Ethereum blockchain}: We run a full Erigon Ethereum archive node to acquire further information on code account creators and cryptoasset balances.
	
	\item \textbf{Ethereum Name Service (ENS)}: Taking advantage of Snapshot's data structure, all DAO spaces are linked to their corresponding ENS domains\footnote{Refer to \url{https://docs.snapshot.org/user-guides/spaces/space-roles}}, which reveal information about addresses associated with a DAO and the roles of users in charge.
	
	\item \textbf{The Graph}: This platform makes DLT data accessible through GraphQL and supplies additional information about ENS domains and their sub-domains\footnote{\url{https://thegraph.com/hosted-service/subgraph/ensdomains/ens}}, which we are able to utilize and interlink.
	
\end{itemize}

\subsection{Contribution roles}\label{sec:Annex_Contributors}

We extract contributions with different roles using the following sources:

\begin{itemize}
	
	\item \textit{Owner}: we retrieve the domain owner address from the ENS service of the Ethereum node and The Graph.
	
	\item \textit{Administrator}: we retrieve the addresses of administrators from Snapshot.
	
	\item \textit{Developer}: from Snapshot we obtain all contract account (CA) addresses associated with a certain space. If a CA represents a governance token, we keep it only if it can unambiguously be associated with that space\footnote{Some spaces might utilize external governance tokens or prominent tokens, like the stablecoin USDT, for voting.}. Then we retrieve the addresses of contract creators from executed on-chain \emph{create} transactions. Adopting a conservative approach, we only consider direct creations by users initiated via their externally owned accounts (EOAs). We discern indirect creations via contract accounts (CAs).

\end{itemize}

\subsection{Total Value Locked}\label{sec:tvl}

We use the Total Value Locked (TVL) as the benchmark of the financial value of a DAO, and therefore as an indicator of its relevance.
We collect the TVL by scraping available dApps from the DeFiLlama API\footnote{\url{https://api.llama.fi/protocols}}.
We focus on those with TVL available on Ethereum, and exclude centralized  entities such as centralized exchanges (CEX). 

To link the TVL data to snapshot DAO spaces, we directly included all scraped protocols that had governance identifiers in Snapshot. For those without such identifiers, we employed a name-matching approach based on Jaccard similarity.
We accomplished this by utilizing an algorithm that compares names and identifier strings between Snapshot and DeFiLlama entries. We introduced a threshold of \num{0.5} for the Jaccard similarity of the names and \num{0.6} for the identifiers to link similar entries. To ensure consistency, we conducted a manual check of the entries and removed those with uncertain matches.

For the remaining matches that passed both the automatic and manual checks, we collected the TVL on Ethereum at the latest available time point in 2022. In cases where DeFi protocols had multiple versions, resulting in a \textit{1:n} relationship between Snapshot and DeFiLlama, we aggregated all versions of the same snapshot identifier to establish a \textit{1:1} relationship. 

\subsection{Data cleaning procedure}\label{sec:Annex_Cleaning}

We now provide additional details on the data cleaning procedure described in Section~\ref{sec:raw}. 
To identify DAO spaces with mature governance structures, we computed four features for each space: the total number of proposals; the number of proposals with more than ten votes; the number of users following a space; and the TVL obtained from DeFiLama.
We then applied the following cleaning steps: first, we removed DAOs with TVL smaller than \num{100}k USD and selected the set of DAOs in the first 500 positions for at least one of the aforementioned measures. 

Next, we removed \ProposalNotFinal/ non-final proposals with \textit{pending} or \textit{invalid} status and only took into account those with positive \textit{final} status. We also restricted our dataset to proposals using the \textit{single-choice} voting type, which reflects the majority of proposals (\ProposalSingleCoicePct/\%). In a single-choice vote, users can choose only one option, which simplifies the evaluation of voting outcomes. Finally, we verify the consistency of the scoring information of each proposal, i.e., we measure if the sum of individual votes and aggregate Snapshot data correspond and remove \ProposalChoiceScoreNotMatch/ inconsistent proposals.

\subsection{Voting strategies}\label{sec:strategies}

There are several ways to enable voting rights on proposals. We denote the methods to determine the power of a vote as \emph{strategies}, following the terminology of Snapshot.

The Snapshot \emph{raw} data set introduced in Section \ref{sec:raw} consists of \VotesRaw/ votes on \ProposalsRaw/ proposals. In these proposals, we identified \StrategiesN/ distinct voting strategies, some of which are specific to individual DAOs. The most prominent one is \emph{erc20-balance-of} ($f^{erc20}$), which defines the voting power to be proportional to the balance of a specific token, typically the DAO governance token. It is used in \StrategiesErcPct/\% of proposals. 
Similarly, the strategy \textit{erc721} ($f^{erc721}$) computes voting power proportionally to the amount of some specific NFTs held by each user and 
\textit{eth-balance} from the ETH balance. Also, the \textit{ticket} strategy is relevant, implementing the ``one person, one vote'' approach. We note that strategy functions can be implemented arbitrarily and be more complex and also exploit holdings of other cryptoassets such as stablecoins (USDT) or governance tokens of other DAOs.
In Table \ref{tab:strategies}, we show the top 20 strategies by number of proposals that use them. Note that strategies can vary significantly, e.g., from using tokens of multiple chains to being tailored to a specific DAO. 
The retrieval of this information necessitates a substantial investment in blockchain infrastructure and ongoing maintenance, concurrently making more complicated for outsiders to validate the correctness.

\begin{table}[t]
	\begin{tabular*}{\columnwidth}{@{\extracolsep{\fill}}lrr}
	\toprule
	{Name} & {Number of proposals} & {Percentage} \\
	\midrule
	erc20-balance-of & \num{24359} & 53.85 \\
	delegation & \num{5815} & 12.86 \\
	multichain & \num{3842} & 8.49 \\
	erc721 & \num{3806} & 8.41 \\
	ticket & \num{3543} & 7.83 \\
	contract-call & \num{3514} & 7.77 \\
	erc721-with-multiplier & \num{2549} & 5.64 \\
	decentraland-estate-size & \num{2260} & 5.0 \\
	pancake & \num{2231} & 4.93 \\
	cake & \num{1350} & 2.98 \\
	erc20-balance-of-delegation & \num{1111} & 2.46 \\
	erc20-with-balance & \num{700} & 1.55 \\
	uniswap & \num{699} & 1.55 \\
	balance-of-with-min & \num{660} & 1.46 \\
	pagination & \num{637} & 1.41 \\
	eth-balance & \num{625} & 1.38 \\
	whitelist & \num{566} & 1.25 \\
	sushiswap & \num{460} & 1.02 \\
	erc1155-balance-of & \num{447} & 0.99 \\
	masterchef-pool-balance & \num{377} & 0.83 \\
	\bottomrule
\end{tabular*}
	\vspace{0.2cm}
	\caption{\textbf{Top 20 Strategies by number of proposals.} Strategies are used to determine the voting power of votes. We list most frequently used ones by number and percentage of proposals.}
	\label{tab:strategies}
\end{table}


\subsection{Data validation procedure}\label{sec:tokens}

\begin{longtable}{m{0.25\linewidth} m{0.6\linewidth} m{0.069\linewidth} m{0.071\linewidth}}
	\toprule
	DAO space & Token & Error & Count \\ 
	\midrule
	bgansv2.eth & 0x31385d3520bced94f77aae104b406994d8f2168c & {[}1] & 3003 \\
	theopendao.eth & 0x3b484b82567a09e2588a13d54d032153f0c0aee0 & {[}1] & 2312 \\
	purrnelopes\/-countryclub.eth & 0x9759226b2f8ddeff81583e244ef3bd13aaa7e4a1 & {[}1] & 2168 \\
	goopsnapshot.eth & 0x15a2d6c2b4b9903c27f50cb8b32160ab17f186e2 & {[}1] & 1677 \\
	9x9x9.eth & 0x5219c2f6f8ed1e76c937ed1269eda2658ba3c721 & {[}1] & 1296 \\
	cryptohoots.eth & 0x5754f44bc96f9f0fe1a568253452a3f40f5e9f59 & {[}1] & 1168 \\
	pandaparadise.eth & 0x24998f0a028d197413ef57c7810f7a5ef8b9fa55 & {[}1] & 202 \\
	cryptohoots.eth & 0x196f7e9c5769fc777909a3f1b9bd65959f3f64fb & {[}1] & 102 \\
	gfanchan11.eth & 0xa0b86991c6218b36c1d19d4a2e9eb0ce3606eb48 & {[}4] & 14 \\
	shibvinci.eth & 0xe9615071341c6f0392a5dfde1645ad01b810cb43 & {[}3] & 9 \\
	gfanchan11.eth & 0x6b175474e89094c44da98b954eedeac495271d0f & {[}1] & 7 \\
	ilvgov.eth & 0xaebd9bd588f044cbdec8f3cf1e80277a7a52dc69 & {[}2] & 4 \\
	gfanchan11.eth & 0xdac17f958d2ee523a2206206994597c13d831ec7 & {[}4] & 3 \\
	\bottomrule
	\caption{\textbf{Report of DAO spaces and patterns of mismatches.} This table contains the DAO spaces with inconsistent values between Snapshot and on-chain data. Patterns of errors can be found and categorized in classes.}
	\label{votesUnsolved}
\end{longtable}

We aim to validate the voting power $w$ of the \emph{cleaned} data set with on-chain data wherever possible. Therefore, we focus on the on-chain Ethereum blockchain and on proposals covered by the selected strategies $F' \subseteq \mathbb{P}( \{f^{erc20}, f^{erc721}, f^{eth} \})$\footnote{Note, however, that in principle the same approach can be applied to multiple chains to retrieve additional voting balances.}. The reasoning for the restriction to $F'$ is that strategies must be re-implemented manually, and $F'$ contains the three most widely adopted strategies (which define the voting power to be proportional to the balance of a specific token, typically the DAO governance token or to the ether balance).

We start by coding the aforementioned strategies, i.e., we gather all asset holdings of the voters and recompute their voting power from on-chain Ethereum data.  
We apply the strategy information given in the Snapshot data. For governance tokens, this includes the token address as well as the decimal, i.e., the decimal floating point value necessary to convert on-chain data to a floating point number. That is, a token balance is reported on-chain as an integer and must be converted to its actual value. For instance, an illustrative token TKN can be reported on-chain with value 10,000,000 but its actual value is 10 because its decimal is 6. 
If no decimal is given, we assume it to be zero. 

When we compare the queried voting power to the Snapshot data, discrepancies arise. A closer inspection revealed that some Snapshot proposals report incorrect values for the decimal value, e.g., there is no correspondence with the value reported in the smart contract of the token, or there is no "BalanceOf" function and therefore it is not possible to convert the on-chain value to the floating point.
To avoid considering rounding errors as mismatches, we introduce a tolerance interval $\Delta\epsilon = 10^{-3}$ and consider the on-chain and off-chain values as equal if the difference is smaller than the threshold. 
We compare the voting power computed with that reported by Snapshot, and found high consistency: with this approach, we could verify that \VotesConsistency/ (\VotesConsistencyPct/\%) of \VotesToValidate/
Ethereum Snapshot weights are correct.
We documented, however, larger deviations for some DAO spaces. 
The list in Table~\ref{votesUnsolved} reports the spaces and the number of mismatches for each space. We find recurring patterns of errors, which we can categorize and also provide solutions to get also consistent values:

\begin{itemize}
	\item Solution [1]: Instead of the decimal value given by the Snapshot strategy, we apply the decimal contained in the token smart contract. If none is given, we consider the decimal equal to \num{0}. 
	\item Solution [2]: We repeat the approach of Solution [1] but consider the decimal equal to \num{18}. 18 is the conversion factor from wei to ether, and some currencies adopted it as well.
	\item Solution [3]: We apply the $\log_{10}$-function for both Snapshot and on-chain data before comparing them because the on-chain data reported as integers are very large, and therefore the deviation exceeds $\Delta\epsilon$, even though the actual values after the conversion would be small.
	\item Solution [4]: We substitute the decimal with \num{18} to convert the floating value.
\end{itemize}

These DAO spaces leave doubts about the reliability of the reported values. 
Thus, we selected only  \SpacesValidated/ out of \SpacesToValidate/  DAOs where all votes have been fully validated.
The resulting \textit{validated dataset}, used to conduct the analyses in Section \ref{sec:shifts},
comprises \VotersValidated/ users, who voted \VotesValidated/ times on \ProposalsValidated/ proposals of \SpacesValidated/ spaces. 
These include \ContrVotesValidated/ contributor votes out of \ContributionValidated/ contributions.

\subsection{Snapshot activity}\label{sec:Annex_Progress}

\begin{figure}[t]
	\includegraphics[width=0.95\columnwidth]{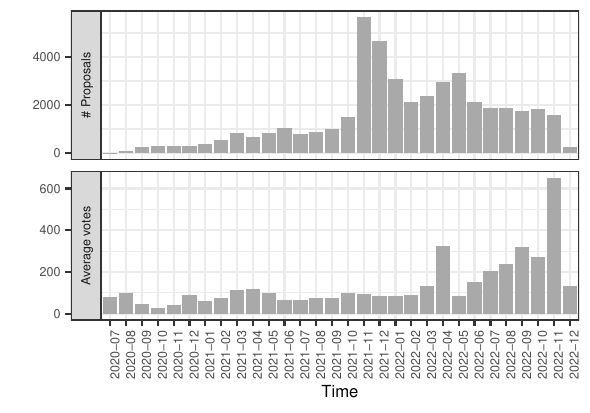}
	\caption{\textbf{Evolution of Snapshot activity in time}. Number of proposals created each month on the Snapshot platform (top) and the average votes per proposal (bottom). The panel on top shows that the number of proposals created monthly on Snapshot has grown significantly in 2021, with a peak of more than 5,000 proposals executed in November 2021. The bottom Panel shows that the average number of votes cast per proposal was around 100 until the beginning of 2022. Since then, this number is followed an increasing trend.}
	\label{Proposals_AvgVotes_YM}
\end{figure}

The voting platform Snapshot gained popularity in recent years, and especially after 2020.
To give an overview, Figure~\ref{Proposals_AvgVotes_YM} shows the evolution over time of the monthly number of proposals and the average votes from July 2020 to December 2022. We observe a trend of increasing proposals and average votes, with peaks in November 2021 for the former and in November 2022 for the latter.

\newpage 

\section{Influence of Contributors on DAO Governance}\label{sec:Annex_Influence}

\setcounter{figure}{0}
\renewcommand\thefigure{\thesection.\arabic{figure}}    
\setcounter{table}{0}
\renewcommand\thetable{\thesection.\arabic{table}}    

We provide more details on the involvement of individual DAOs (Section \ref{sec:DAOinf}) and on self-decisions (Section \ref{sec:Annex_SelfDecision}). First, we provide additional statistics on the contributor involvement and second, we measure to what extent other-space contributors overrule the same-space contributors in making decisions.

\subsection{Contributor involvement}\label{sec:DAOinf}

\begin{figure}[t]
	\centering
	\includegraphics[width=1\columnwidth]{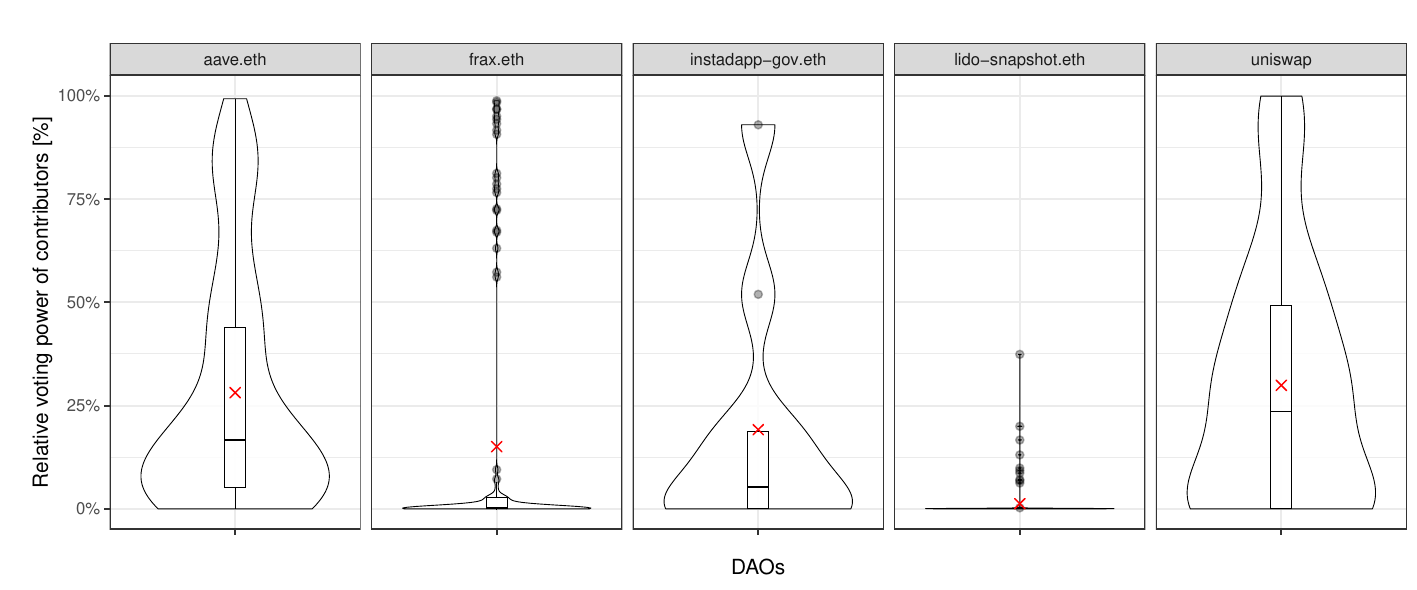}
	\caption{\textbf{Contributors' involvement in top-TVL DAOs}. Box- and violin plots of relative voting power of contributors across top TVL DAOs. We find the mean values (\textcolor{red}{\text{\sffamily X}}) above the median,  for some proposals but outliers even above \num{50}\% relative voting power. Thus, also in high TVL DAOs, contributors had majorities in some proposals.}
	\label{fig:DAO_assoc_vp_violin}
\end{figure}

We complement the analysis in Section \ref{sec:involement} by providing additional statistics on the top spaces by TVL and their contributor involvement.
Table~\ref{tab:contrInv50} lists DAOs with contributor involvement above \num{50}\% and additional statistical measures. Figure~\ref{fig:DAO_assoc_vp_violin} shows the distributions across proposals for the top five spaces by TVL. In all these plots, the median is less than the average, but we also find outliers even above \num{50}\%, meaning that contributors had majorities in certain proposals.

\begin{table}
	\centering
	\scriptsize
	\begin{tabular*}{\columnwidth}{@{\extracolsep{\fill}}lrrrrrr}
\toprule
{DAOs} & {\# Proposals} & \multicolumn{5}{c}{Contributor Involvement} \\
{} & {} & {Mean} & {Max} & {Std} & {Min} & {Median} \\
\midrule
levidao.eth & \num{185} & 100.0\% & 100.0\% & 0.0\% & 100.0\% & 100.0\% \\
cryptodigger.eth & \num{29} & 100.0\% & 100.0\% & 0.0\% & 100.0\% & 100.0\% \\
zingan.eth & \num{4} & 100.0\% & 100.0\% & 0.0\% & 100.0\% & 100.0\% \\
651792.eth & \num{22} & 100.0\% & 100.0\% & 0.0\% & 100.0\% & 100.0\% \\
strikeorg.eth & \num{1} & 100.0\% & 100.0\% & -\% & 100.0\% & 100.0\% \\
alongdomainisacheapdomain.eth & \num{23} & 100.0\% & 100.0\% & 0.0\% & 100.0\% & 100.0\% \\
oligamy.eth & \num{2} & 100.0\% & 100.0\% & 0.0\% & 100.0\% & 100.0\% \\
atato.eth & \num{22} & 100.0\% & 100.0\% & 0.0\% & 100.0\% & 100.0\% \\
loganzc.eth & \num{28} & 100.0\% & 100.0\% & 0.0\% & 100.0\% & 100.0\% \\
test4.shot.eth & \num{28} & 50.2\% & 100.0\% & 42.0\% & 0.0\% & 50.0\% \\
longbtc20090103.eth & \num{20} & 50.23\% & 100.0\% & 23.02\% & 30.77\% & 41.43\% \\
castledao.eth & \num{3} & 50.26\% & 56.98\% & 6.04\% & 45.28\% & 48.51\% \\
mcv.eth & \num{59} & 50.48\% & 95.01\% & 16.08\% & 0.0\% & 47.65\% \\
comp-vote.eth & \num{5} & 51.15\% & 100.0\% & 36.55\% & 0.02\% & 49.62\% \\
coinpirates.eth & \num{2} & 51.23\% & 70.08\% & 26.65\% & 32.38\% & 51.23\% \\
seedifyfund.eth & \num{22} & 51.3\% & 100.0\% & 42.89\% & 0.0\% & 28.57\% \\
apeswap-finance.eth & \num{24} & 52.02\% & 99.37\% & 40.76\% & 0.0\% & 62.19\% \\
5021314.eth & \num{17} & 52.94\% & 100.0\% & 43.79\% & 0.0\% & 66.67\% \\
sarcophagus-ambassadors.eth & \num{41} & 54.81\% & 100.0\% & 28.86\% & 0.13\% & 59.03\% \\
rabbitholes.eth & \num{6} & 56.95\% & 100.0\% & 45.48\% & 0.0\% & 71.27\% \\
lxdao.eth & \num{37} & 57.5\% & 100.0\% & 32.53\% & 18.18\% & 42.86\% \\
ffdao.eth & \num{25} & 57.69\% & 86.59\% & 17.32\% & 19.03\% & 58.99\% \\
usdaogov.eth & \num{21} & 57.9\% & 100.0\% & 43.55\% & 0.0\% & 77.78\% \\
sdbal.eth & \num{31} & 58.02\% & 100.0\% & 50.12\% & 0.0\% & 100.0\% \\
layer2finance.eth & \num{2} & 58.73\% & 84.87\% & 36.96\% & 32.59\% & 58.73\% \\
meebitsdaopool.szns.eth & \num{22} & 59.02\% & 100.0\% & 41.14\% & 0.0\% & 74.94\% \\
metrox.eth & \num{34} & 59.54\% & 100.0\% & 46.08\% & 0.0\% & 99.93\% \\
ginoct.eth & \num{64} & 60.99\% & 100.0\% & 47.5\% & 0.0\% & 100.0\% \\
weenus & \num{44} & 61.61\% & 100.0\% & 47.63\% & 0.0\% & 100.0\% \\
dorg.eth & \num{161} & 62.81\% & 100.0\% & 27.54\% & 0.0\% & 67.56\% \\
al409.eth & \num{17} & 65.11\% & 100.0\% & 36.51\% & 0.0\% & 66.67\% \\
dakshow.eth & \num{25} & 66.67\% & 100.0\% & 43.03\% & 0.0\% & 100.0\% \\
alexec.eth & \num{3} & 66.69\% & 100.0\% & 57.7\% & 0.06\% & 100.0\% \\
wintersun.eth & \num{11} & 68.79\% & 76.04\% & 13.53\% & 28.55\% & 73.37\% \\
cryengine.eth & \num{13} & 69.01\% & 85.03\% & 17.13\% & 31.0\% & 75.4\% \\
pharo.eth & \num{1} & 69.37\% & 69.37\% & -\% & 69.37\% & 69.37\% \\
insuretoken.eth & \num{36} & 69.4\% & 100.0\% & 46.69\% & 0.0\% & 100.0\% \\
wdefi.eth & \num{23} & 70.94\% & 100.0\% & 42.61\% & 0.0\% & 99.54\% \\
thanku.eth & \num{21} & 71.43\% & 100.0\% & 46.29\% & 0.0\% & 100.0\% \\
primerating.eth & \num{249} & 72.06\% & 100.0\% & 34.75\% & 0.0\% & 100.0\% \\
cvx.eth & \num{265} & 72.44\% & 99.58\% & 26.34\% & 0.0\% & 77.57\% \\
testbsw.eth & \num{17} & 76.4\% & 100.0\% & 43.68\% & 0.0\% & 100.0\% \\
venus-xvs.eth & \num{17} & 77.47\% & 99.88\% & 36.07\% & 0.0\% & 91.47\% \\
lemu.dcl.eth & \num{42} & 78.16\% & 100.0\% & 41.4\% & 0.0\% & 100.0\% \\
huangwenchao.eth & \num{72} & 78.24\% & 100.0\% & 37.7\% & 0.0\% & 100.0\% \\
melson.eth & \num{14} & 78.82\% & 100.0\% & 19.84\% & 27.28\% & 82.33\% \\
cabindao.eth & \num{4} & 79.53\% & 97.16\% & 20.28\% & 56.86\% & 82.05\% \\
grantsdao.eth & \num{36} & 80.42\% & 100.0\% & 20.64\% & 33.33\% & 77.5\% \\
polywrap.eth & \num{132} & 80.58\% & 100.0\% & 20.03\% & 0.0\% & 83.57\% \\
cgpool.eth & \num{1} & 81.04\% & 81.04\% & -\% & 81.04\% & 81.04\% \\
4.spaceshot.eth & \num{9} & 81.19\% & 100.0\% & 34.68\% & 0.24\% & 100.0\% \\
xunfa.eth & \num{35} & 81.86\% & 100.0\% & 5.91\% & 79.83\% & 79.83\% \\
tommyg.eth & \num{14} & 83.33\% & 100.0\% & 36.4\% & 0.0\% & 100.0\% \\
retokendao.eth & \num{36} & 84.46\% & 100.0\% & 10.09\% & 54.55\% & 85.43\% \\
frami.eth & \num{206} & 85.17\% & 100.0\% & 34.98\% & 0.0\% & 100.0\% \\
adaocompany.eth & \num{20} & 85.85\% & 100.0\% & 32.06\% & 0.0\% & 100.0\% \\
3.spaceshot.eth & \num{69} & 86.22\% & 100.0\% & 23.12\% & 0.0\% & 100.0\% \\
mycontext.eth & \num{9} & 87.04\% & 100.0\% & 26.06\% & 33.33\% & 100.0\% \\
* & \num{15} & 87.78\% & 100.0\% & 28.5\% & 0.0\% & 100.0\% \\
btc1.eth & \num{2} & 89.49\% & 100.0\% & 14.87\% & 78.97\% & 89.49\% \\
szns.shean.eth & \num{61} & 91.15\% & 100.0\% & 27.93\% & 0.0\% & 100.0\% \\
bridgeswap.eth & \num{25} & 91.73\% & 98.7\% & 5.11\% & 83.9\% & 94.69\% \\
oracles.opiumprotocol.eth & \num{25} & 94.77\% & 100.0\% & 18.74\% & 8.77\% & 100.0\% \\
ecashxec.eth & \num{1} & 95.82\% & 95.82\% & -\% & 95.82\% & 95.82\% \\
halodao-kovan.eth & \num{41} & 97.56\% & 100.0\% & 15.62\% & 0.0\% & 100.0\% \\
alexjyoung.eth & \num{24} & 98.75\% & 100.0\% & 6.11\% & 70.05\% & 100.0\% \\
\bottomrule
\end{tabular*}

	\vspace{0.2cm}
	\caption{\textbf{Statistics of contributor voting power.} Additional statistics on voting power for the \SpacesNVPmoreFifty/ DAOs with mean contributor involvement above \num{50}\%.}
	\small\textsuperscript{*} DAO name contains character that cannot be displayed.
	\label{tab:contrInv50}
\end{table}

\subsection{Contributor self-decisions}\label{sec:Annex_SelfDecision}

\subsubsection{Decisions by other-space contributors}

To provide a broader picture, we complement the measure of \textit{contributor self-decisions}, introduced in Section~\ref{sec:influence}, with the \textit{contributor other-decisions}, that is, we measure to what extent other-space contributors overrule the same-space contributors in making decisions.
We define three joint conditions for proposals to be \textit{decisive other-votes}. The first two conditions are inspired by \textit{decisive self-votes} applied to votes of other-space contributors, which we denote as ``other-contributor votes''. First, other-contributor votes have a relative majority within the voting power for the decision. Second, we consider also the second-ranked option and select those proposals where the weight of other-space contributors in the decision is higher than the weight of the second-ranked option. Third, we take into account only proposals where same-space contributors are involved in the voting for the second ranked option.

Formally, we define the set of \textit{decisive other votes} $V^{p}_{DO} = V^p_{\hat{o}1} \cap V^p_{OS}$ and also the complement set $V^p_{CO} = V^p_{\hat{o}1} \, \backslash \, V^{p}_{DO}$; we identify the fractions of relative voting power for decisive other votes (\ref{eq:relVpOV}), for the complement set (\ref{eq:relVpCO}) and the same-space contributors for the second option  $V^{p}_{SS2} = V^p_{\hat{o}2} \cap V^p_{SS}$ as 

\vspace{0.15cm}

\noindent\begin{minipage}{.3\linewidth}
	\begin{equation}
	\tilde{w}^{p}_{DO} = \sum_{v_i \in V^p_{DO}} \tilde{w_i}, \label{eq:relVpOV}
	\end{equation}
\end{minipage}
\begin{minipage}{.3\linewidth}
	\begin{equation}
	\tilde{w}^{p}_{CO} = \sum_{v_i \in V^p_{CO}} \tilde{w_i}, \label{eq:relVpCO}
	\end{equation}
\end{minipage}	
\begin{minipage}{.3\linewidth}
	\begin{equation}
	\tilde{w}^{p}_{SS2} = \sum_{v_i \in V^p_{SS2}} \tilde{w_i}, \label{eq:relVpSS2}
	\end{equation}
\end{minipage}

\vspace{0.25cm}

We consider proposals to be decided by other-contributors, when all three conditions are fulfilled:

\vspace{-0.15cm}

\begin{align}
[
 \,
(\tilde{w}^{p}_{DO}  > \tilde{w}^{p}_{CO} )
\land
(\tilde{w}^{p}_{DO} > \tilde{w}^{p}_{SS2} )
\land
(\tilde{w}^{p}_{SS2} > 0)
\,
],	
\end{align}

\begin{figure}[!]
	\includegraphics[width=0.95\columnwidth]{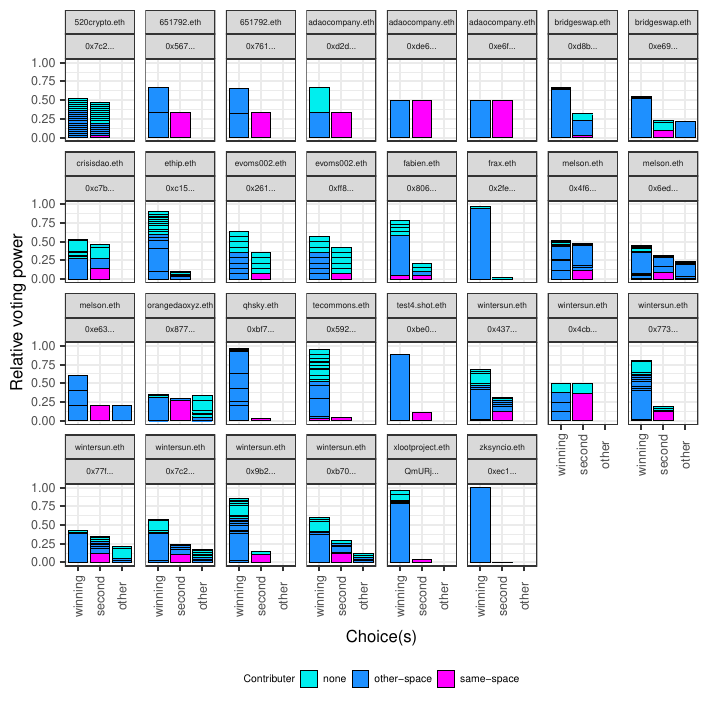}
	\caption{\textbf{Decisions of other-space against same-space contributors}. In 30 proposals, other-space contributors  (\protect\tikz[baseline=0ex]\protect\draw[fill={rgb,255:red,30;green,144;blue,255},line width=1pt]  (0,0) rectangle ++(0.2,0.2);) 
	had the relative majority in the winning choice and 
	overruled the same-space contributors (\protect\tikz[baseline=0ex]\protect\draw[fill={rgb,255:red,255;green,0;blue,255},line width=1pt]  (0,0) rectangle ++(0.2,0.2);), voting for the second choice.	
}
	\label{DAO_selfAgainstNone}
\end{figure}

%
In rare cases, we find that the outcome of a proposal has been dominated by contributors of other spaces over the self-contributors of the proposal's space. We show the relative voting power for these proposals in Figure~\ref{DAO_selfAgainstNone}, by a stacked bar plot of votes on different choices. Each proposal is illustrated by three bars, respectively corresponding to the relative voting power of the winning choice, the second choice by voting power and all other choices aggregated.

\section{Network Analysis}\label{appendix_network}

\setcounter{figure}{0}
\renewcommand\thefigure{\thesection.\arabic{figure}}    
\setcounter{table}{0}
\renewcommand\thetable{\thesection.\arabic{table}}    

\begin{table}[t]\centering
	\centering

\begin{tabular*}{\columnwidth}{@{\extracolsep{\fill}}lrrrr}
    \toprule
    Network &  $G_{AA}$ &  $G_{AW}$ & $G_{TA}$ &  $G_{TW}$\\
    \hline
        Daos &
        All & All & Top-100 & Top-100 \\
                Votes &  All &  Winning & All & Winning \\
    \midrule
    \midrule
            Num Nodes & \num{104863} & \num{75879} & \num{20401} & \num{14494} \\
            Num Edges & \num{739813062} & \num{107374710} & \num{19917792} & \num{6045065} \\
        Max Degree & \num{57304} & \num{31664} &  \num{14456} & \num{9039} \\
        Avg. Degree & \num{14110.09} & \num{2830.16} &  \num{1952.63} & \num{834.15} \\
        \midrule
        \midrule
        Contr. Nodes &  1.29\% & 1.45\% &  3.25\% & 4.5\% \\
        Contr. Edges & 1.61\% & 1.76\% &  3.4\% & 8.0\% \\
        Contr. Max. Degree   & \num{57303} & \num{24346} &  \num{14196} & \num{7755} \\
        Contr. Avg. Degree   & \num{8795.3} & \num{3434.12} &  \num{2079.8} & \num{1490.86} \\
        \midrule
        \midrule
        Giant Component & 98.1\% & 98.8\% &  99.9\% &          99.4\% \\
        Assortativity & 0.02 & 0.05 & 0.11 & 0.07 \\
        Avg. Path Length &      2.24 & 2.58 &       2.21 &       2.37 \\
        Diameter &      32 & 21 &              7 &          9 \\
        Clustering &      1.55 & 1.59 &              1.99 &          2.06 \\
        Small Worldliness &  x & 3.85 &  4.31 &  6.41 \\
        Louvain Communities & x & 330 & 28 & 12 \\
        Largest Community & x & 42\% & 43\% & 34\% \\
    \bottomrule
\end{tabular*}

	\vspace{0.2cm}
	\caption{\textbf{Network statistics of four co-voting networks.} The top of the table defines the four networks as a unique combination of two features: DAOs and Votes. \textit{Top-100} DAOs are ranked by total value locked (TVL); \textit{Winning} votes are votes for the choice that ultimately won the majority of voting power. The rest of the table contains network statistics for the four graphs, the central part referring to all nodes and edges, and the bottom parts only to the giant component. \textit{Some statistics could not be computed for the largest network due to memory limitations.}}
	\label{table_network_descriptive_long}
\end{table}
 
We present additional material about network descriptive statistics (\ref{sec:Covoting_statistics}), an analysis of the small-world properties of the networks (\ref{sec:Annex_small_world}), an investigation of the network of contributors (\ref{sec:contributors_network}), additional plots for the centrality and k-core analysis (\ref{sec_appendix_centrality_kcore}), and finally, a sensitivity analysis for the choice of the edge threshold (\ref{sec_appendix_sensitivity_threshold}).

\subsection{Co-voting network statistics}\label{sec:Covoting_statistics}

In Section \ref{sec:network} we built the four co-voting networks $G_{AA}$, $G_{AW}$, $G_{TA}$, $G_{TW}$.  Their exhaustive statistics are presented in Table~\ref{table_network_descriptive_long}. The table is split in four segments, starting on top with the distinction of the networks by combination of their features. Then, it provides descriptive statistics of the network nodes and edges, of the contributor nodes only, and, finally, insights on network measures computed on the giant component.

\subsection{Small-world properties}\label{sec:Annex_small_world}

The co-voting network with all DAOs and all votes ($G_{AA}$) counts \num{104 863} nodes and is about 5 times 
larger than the corresponding network with the top-100 DAOs by TVL ($G_{TA}$). 
Despite the different sizes, the four networks present similar topological features:
a very large giant component (about 99\% of all nodes) with clear small world features. 
Following the procedure in \cite{sinani2008corporate}, we computed
the ``small-worldliness'' coefficient of each graph, comparing the ratio of average path length 
(APL) and clustering coefficient (CC) normalized by the values of those statistics
in a random network (rnd) with the same number of nodes and degree distribution (test):

\begin{align}	
SW = (APL_{test} / APL_{rnd} ) /  ( CC_{test} / CC_{rnd} )
\end{align}

Intuitively, to be a small world, a network needs APL to be similar to that of a
random network, but CC should be much larger, leading to a ``small-worldliness''
coefficient \textit{SW} considerably larger than one. This is indeed the case for all our networks (\textit{SW} between 3.85 and 6.41). The non-random
structure is also confirmed by a small, yet positive assortativity coefficient, indicating a certain degree of homophily.

\subsection{The contributors network}\label{sec:contributors_network}

Users might connect spaces through their contributions. We construct the \textit{contributors network} $G_{SU}$ as a bipartite network of the contribution relation $\mathcal{C}$, using spaces $\mathcal{S}$ and users $\mathcal{U}$ as two node types without taking into account their specific roles $\mathcal{T}$.

\begin{figure}[t]
	\centering
	\includegraphics[width=0.8\columnwidth]{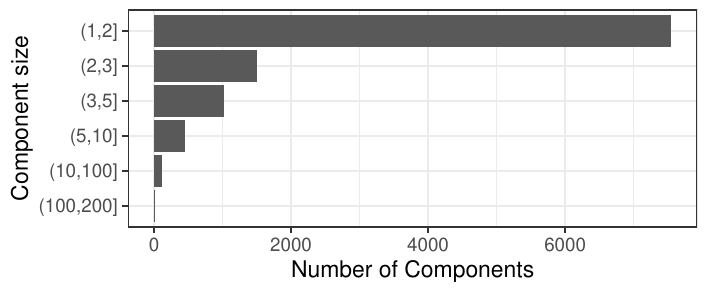}
	\caption{\textbf{Number of connected components of the $G_{CD}$ network}. The network  mainly consists of small components and a larger component with above 100 nodes.}
	\label{ComponentsSizeNum}
\end{figure}

\begin{figure}[h!]
	\centering
	\includegraphics[width=0.8\columnwidth]{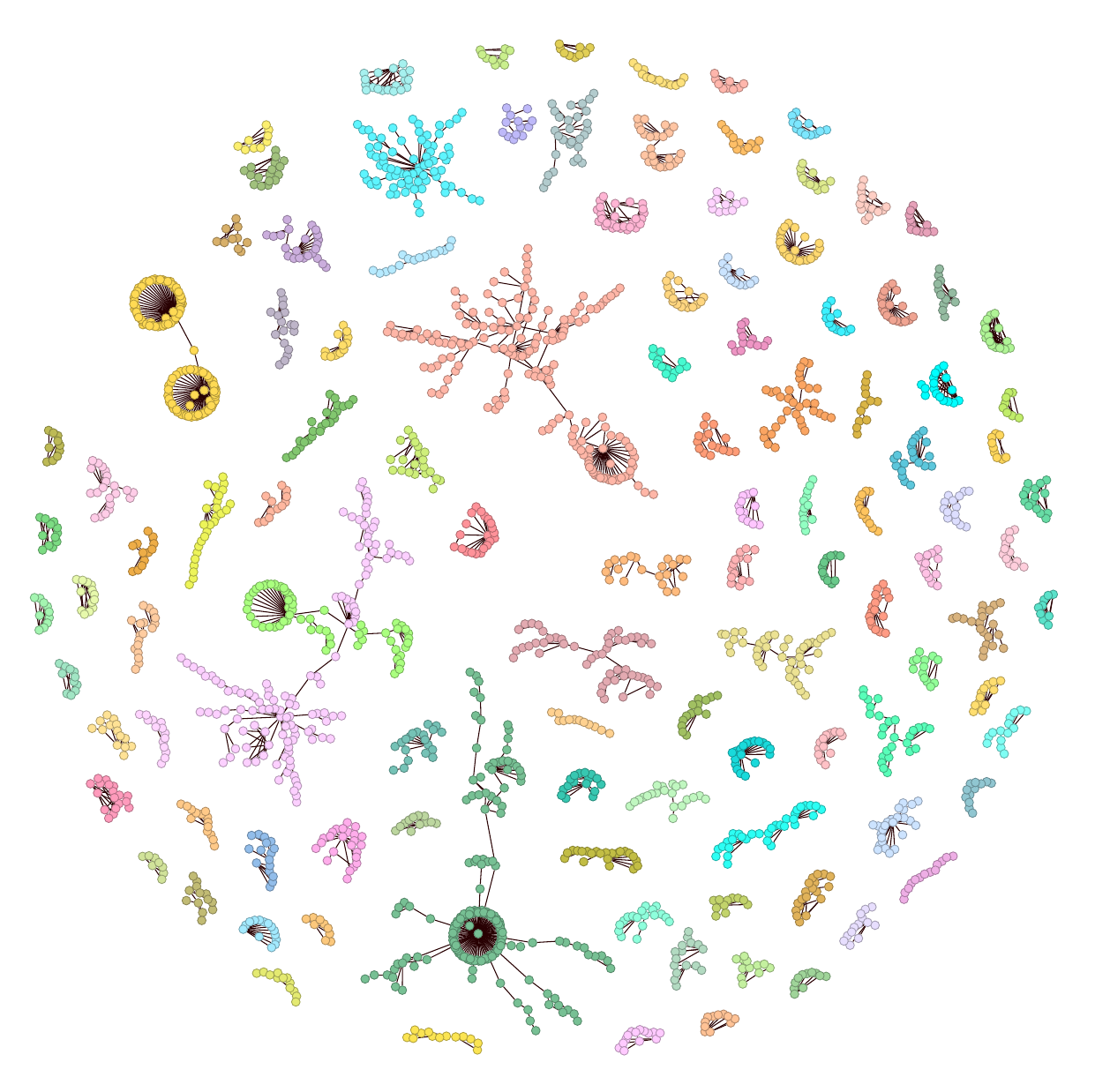}
	\caption{\textbf{Network of DAOs connected trough contributions}. Bipartite network projection of contributors network $G_{SU}$ with more than 10 members. The network illustrates nodes that are linked when they have a common contribution. It appears fragmented, with star-like structures and a view larger bridged DAO communities.}
	\label{DAO_assoc_netw}
\end{figure}

We utilized the raw data set from Section~\ref{sec:raw}, containing \SpacesRaw/ DAO spaces, and extracted the contribution roles with sources defined in Section~\ref{sec:Annex_Contributors}.
Using these sources, we found a total of \SpaceHasContributors/ DAOs \ContributorsN/ addresses. \ContributionRaw/ of these addresses are voters in the raw data, which would strictly interpreted only be identified as contributions. However, we are interested in the most comprehensive network of contributors to DAOs to find all possible relations.
We construct the network $G_{SU}$ linking \SpaceHasContributors/ DAOs to all \ContributorsN/ contribution addresses. 
Thus,
on average, each DAO has \SpaceContributorsAvg/ contributors. This network appears fragmented, as it separates into \ContributorComponents/ small components, mostly with ten or fewer nodes and with rather simple relations between addresses and a few larger ones. In Figure~\ref{ComponentsSizeNum} we plot the distribution of the connected component. 
We limit the network to components size larger than \num{10} and compute the bipartite projection of $G_{SU}$ on DAO spaces. The resulting network in Figure~\ref{DAO_assoc_netw} links DAO nodes when they have contributors in common. We observe several star-like structures, but there also exist larger structures created by contributors bridging multiple DAOs together. 
However, keeping the distribution of Figure~\ref{ComponentsSizeNum} in mind, most of the relation consist only of rather simple relations between DAOs.

\pagebreak

\subsection{Centrality and k-core}\label{sec_appendix_centrality_kcore}

\begin{figure}[t]
	      \centering
	      \includegraphics[width=1\columnwidth]{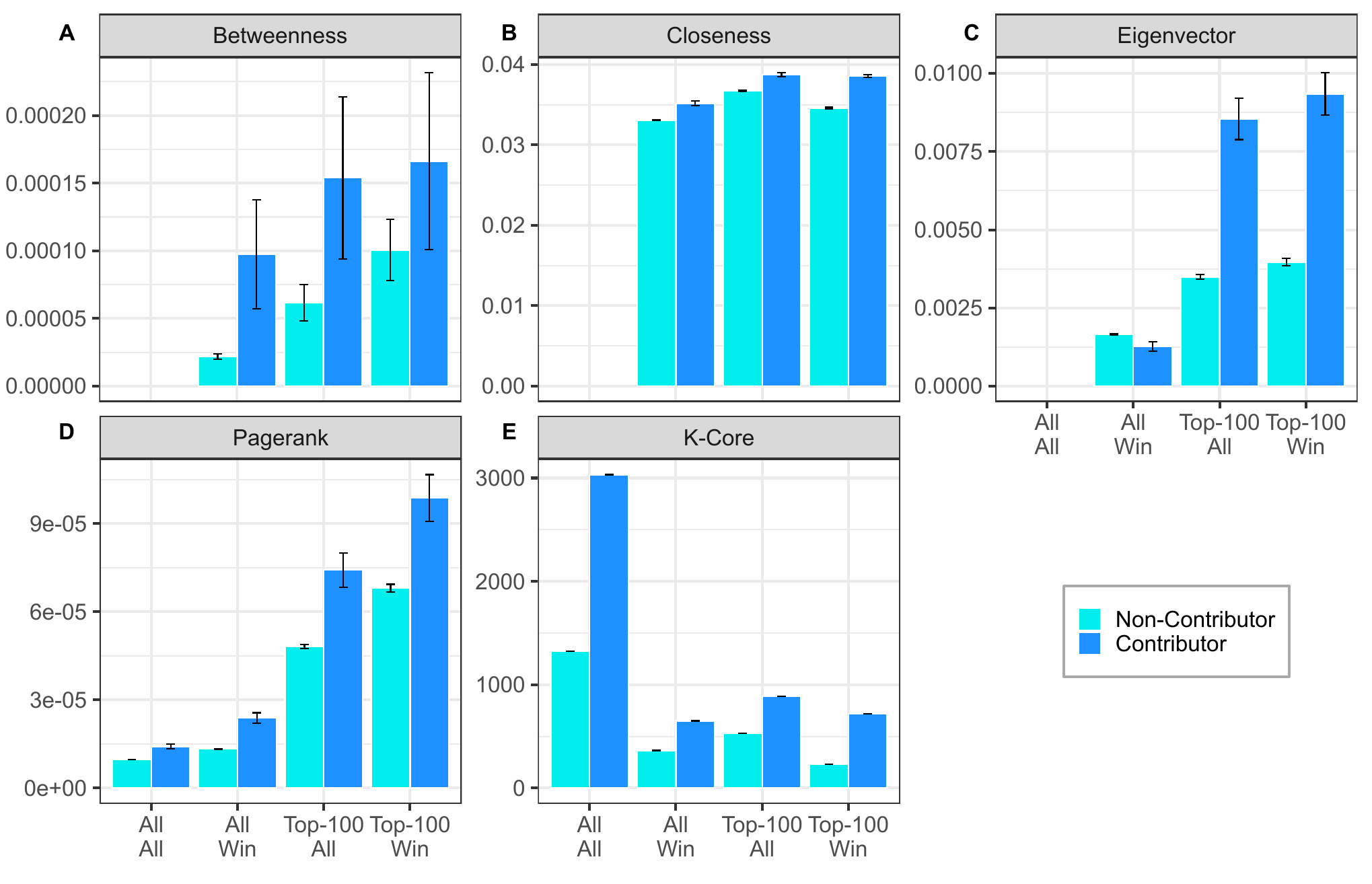}
	      \caption{\textbf{Centrality and k-core statistics in the four co-voting networks}. Contributors tend to have higher centrality and k-core across networks and measures. Overall, it suggests that contributors have influence over the entire network. \textbf{A-D.} Centrality measures (some measures for the largest network could not be computed due to memory limitations). All centrality
	      measures make use of edge weights and are applied on the giant connected component. \textbf{E.} Geometric mean of K-core.
	      Error bars are 95\% confidence intervals of the means.}
	      \label{fig_centrality_all_appendix}
	\end{figure}

We report in Figure \ref{fig_centrality_all_appendix} all the plots of all centrality measures mentioned in the main text. Contributors tend to score higher in all centrality measures, but one (eigenvector centrality in the network with all DAOs and only winning votes).

\begin{figure}[t]
	\centering
	\includegraphics[width=0.9\columnwidth]{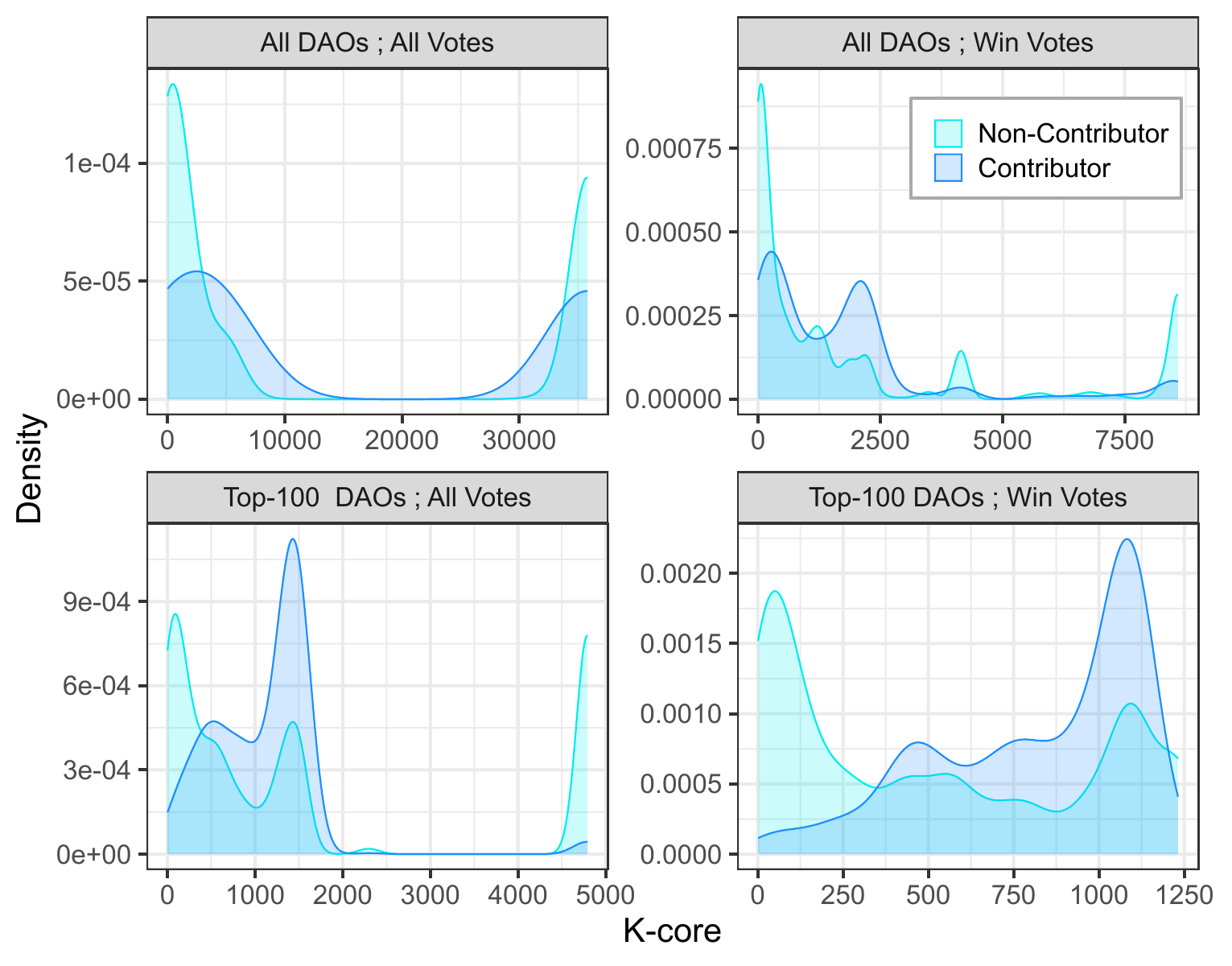}

	\caption{\textbf{K-core density plots for contributor and non-contributor nodes in the four co-voting networks}. Contributors are less frequent in the lower portion of the distribution in all plots. In two plots ($G_{AW}$, $G_{TA}$) an outlier cluster of non-contributor with very high k-core is detected.}
	\label{fig_kcore_density}
\end{figure}

In addition, in Figure \ref{fig_kcore_density}, we report the density plots for the distributions of k-core across all networks. Contributors are less frequent in the portion of the distributions with lower k-core; however, in two networks ($G_{AW}$, $G_{TA}$) an outlier cluster of non-contributors with very high k-core is detected. For this reason, we opted to report the geometric mean in the main text. Econometric tests with the arithmetic means are mixed: contributors have on average significantly higher k-core (t-test $p < 0.001$) in two out of four networks.

\subsection{Edge-weight threshold sensitivity analysis}\label{sec_appendix_sensitivity_threshold}

\begin{figure}
	\centering
	\includegraphics[width=0.99\linewidth]{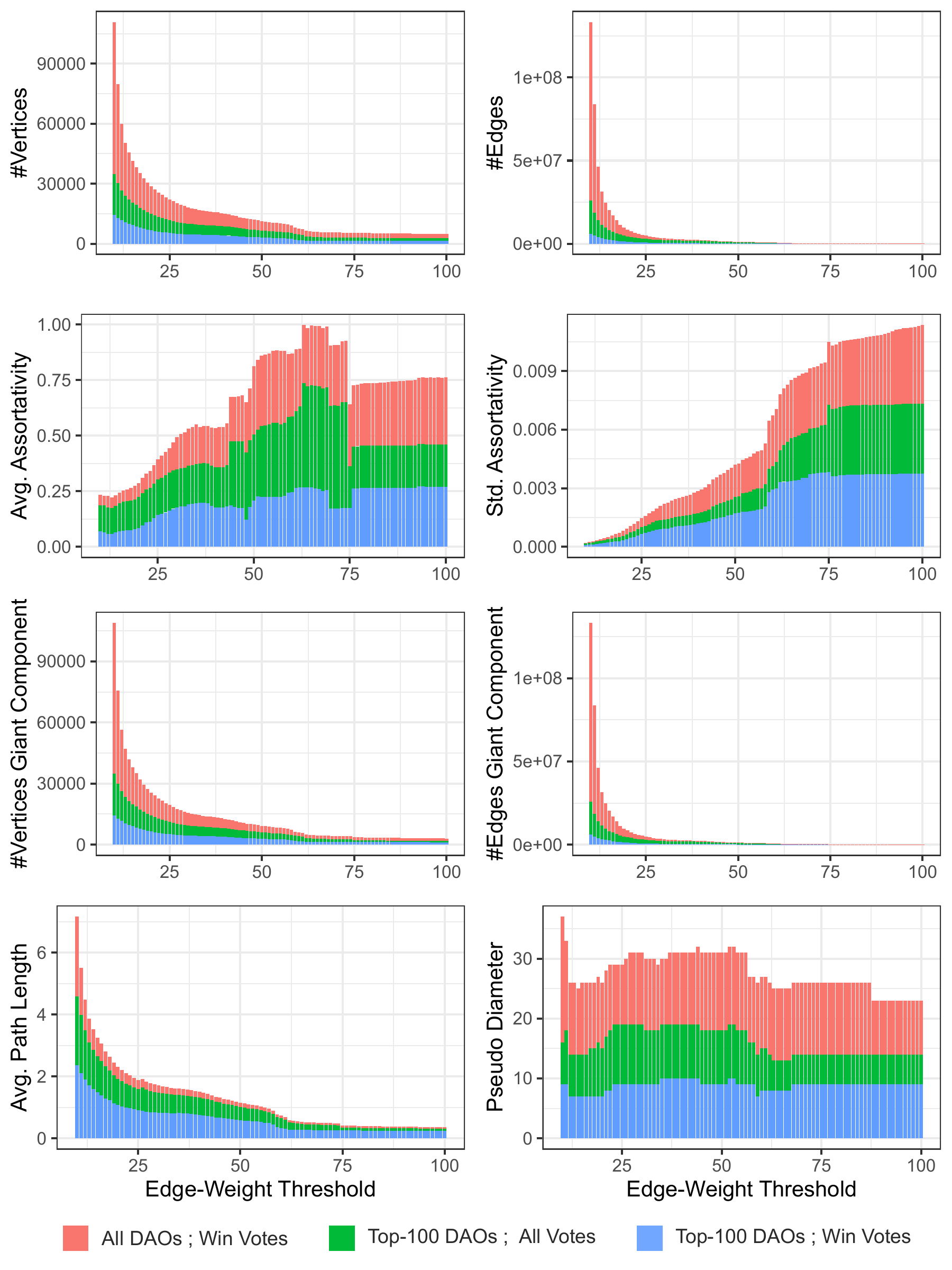}
	\caption{\textbf{Overview of network measures for different edge-weight thresholds.} Two nodes must have co-voted in more than $T$ proposals, where $T$ is the edge-weight threshold.
	The value $T=10$ used in the main text corresponds to the left-most point on the x-axis. All plots are stacked.}
	\label{fig_thresholds_overview}
\end{figure}

In the main text, we report results for networks built with an edge-weight threshold
equal to ten, so that an edge between two nodes is added if and only they have have co-voted in more than 
\textit{T=10} proposals. The value of ten has been conservatively chosen to be low enough to keep as much of the original network structure intact, but at the same time, allowing us to reduce both data noise and the complexity of computation. We tested how the choice of a threshold affected the main network statistics, and the results are reported in Figure \ref{fig_thresholds_overview}. 
For most statistics---i.e., number of edges, number of nodes, average path length---the curve
varies smoothly with $T$; however, assortativity and the pseudo-diameter
seem more sensitive to the choice of the threshold. Further analyses may precisely quantify the impact of the choice of threshold on our results.

\clearpage

\end{document}